\documentclass[
 aip,
 amsmath,amssymb,
 reprint,%
]{revtex4-2}

\newcommand{\ka}{\mathbf{k}}

\newcommand{\ra}{\mathbf{r}}

\newcommand{\R}{\mathbf{R}}

\usepackage{graphicx}
\usepackage{dcolumn}
\usepackage{bm}
\usepackage{algorithm}
\usepackage{algorithmic}
\usepackage{xcolor}
\usepackage{caption}
\usepackage{multirow}
\usepackage{makecell}

\usepackage{hyperref}
\usepackage[capitalise]{cleveref}
\usepackage[utf8]{inputenc}
\usepackage[T1]{fontenc}
\usepackage{mathptmx}
\usepackage{etoolbox}
\usepackage{textcomp}
\usepackage{array}

\newcolumntype{C}[1]{>{\centering\arraybackslash}p{#1}}

\usepackage{xr-hyper} 

\makeatletter
\def\@email#1#2{
 \endgroup
 \patchcmd{\titleblock@produce}
  {\frontmatter@RRAPformat}
  {\frontmatter@RRAPformat{\produce@RRAP{*#1\href{mailto:#2}{#2}}}\frontmatter@RRAPformat}
  {}{}
}

\makeatother
\begin{document}

\preprint{AIP/123-QED}

\title{Efficient All-Electron Periodic Fourier-Transformed Coulomb Method}

\author{Hieu Q. Dinh}
\altaffiliation{These authors contributed equally to this work.}
\affiliation{Department of Chemistry and Chemical Biology, Harvard University, Cambridge, MA, USA}

\author{Adam Rettig}
\altaffiliation{These authors contributed equally to this work.}
\affiliation{Department of Chemistry and Chemical Biology, Harvard University, Cambridge, MA, USA}

\author{Xintian Feng}
\affiliation{Q-Chem, Inc., 6601 Owens Drive, Suite 240, Pleasanton, California 94588, United States}

\author{Joonho Lee}
\email{joonholee@g.harvard.edu}
\affiliation{Department of Chemistry and Chemical Biology, Harvard University, Cambridge, MA, USA}

\date{\today}

\begin{abstract}
    We present an efficient algorithm for the all-electron periodic Coulomb matrix based on the Ewald summation combined with the Fourier-transformed Coulomb method. 
    The short-range contributions involving compact densities are evaluated in real space using the standard Gaussian density fitting method. 
    For the long-range contributions, we introduce an integral-direct planewave density fitting, applicable to both compact and diffuse densities.
    The resulting method achieves orders-of-magnitude speedups for prototypical solid-state systems compared to a closely related approach, the range-separated density fitting method.
    Using dispersion-corrected PBE functional and all-electron Dunning and Karlsruhe basis sets, we apply our method to compute the cohesive energy of the benzene crystal and the adsorption energy of CO on the MgO(001) surface.
    These results are in good agreement with existing literature. 
    Our approach enables efficient Gaussian-based semi-local density functional calculations using dense $\mathbf {k}$-point meshes and traditional molecular Gaussian basis sets.
\end{abstract}

\maketitle
\section{Introduction} \label{sec:introduction}
Kohn-Sham (KS) Density Functional Theory (DFT) \cite{parr1989density} has been established as the primary workhorse for quantum molecular and condensed-phase simulations due to its excellent balance between accuracy and speed.
For condensed-phase applications, the most common approaches to DFT employ planewaves (PWs) as the primary basis set.
With this choice of discretization, the bottleneck is the diagonalization of the Fock matrix (without exact exchange), which scales as $\mathcal O(N_kN^3)$, where $N_k$ is the number of $\mathbf k$-points and $N$ is the system size, without any additional tricks.

For Gaussian-type orbitals (GTOs), by exploiting the sparsity of atomic orbitals, one can reach the same asymptotic scaling. \cite{ochsenfeld2007linear}
Nonetheless, time-to-solution for a similar accuracy has often been shorter with PWs than GTOs for solids. \cite{ulian2013comparison} 
Furthermore, the emergence of linear dependencies in GTOs optimized for molecules (e.g., Dunning, \cite{dunning1989gaussian, woon1993gaussian} Karlsruhe, \cite{weigend2005balanced} Jensen, \cite{jensen2001polarization} etc.) has made the application of GTOs even more challenging, often resulting in the development of specialized basis sets for solids.~\cite{vandevondele2007gaussian,peintinger2013consistent,laun2018consistent,daga2020gaussian,ye2022correlation} 
While the natural capability for all-electron calculations is attractive, these two drawbacks, larger computational costs and linear dependencies, have put GTOs behind PWs.

Some of us have demonstrated that the linear dependencies of GTOs can be adequately handled through canonical orthogonalization, achieving total energies comparable to those of converged PW results using large uncontracted GTO basis sets.~\cite{lee2021approaching} 
This is the same method used for linear dependencies arising in large molecular calculations.~\cite{lowdin1970nonorthogonality, szabo1996modern} 
On the computational front, achieving $\mathcal O(N)$ scaling for building the exchange-correlation (XC) Fock contribution is relatively straightforward using Becke's atom-centered grids.~\cite{becke_multicenter_1988, franchini_becke_2013} 
Calculating the Coulomb matrix presents more delicate considerations, such as the shell-pair sparsity and Coulomb singularity.
In this work, we present a fast algorithm for building the Coulomb matrix such that we can study solid-state materials efficiently with standard molecular GTO basis sets using the periodic boundary condition program in \texttt{Q-Chem},~\cite{epifanovsky2021software} \texttt{QCPBC}.~\cite{lee2021approaching, lee2022faster,rettig2023even} 

Various strategies exist for computing the Coulomb matrix without explicitly generating the expensive two-electron four-center integrals. 
If the McMurchie-Davidson (MD) algorithm \cite{mcmurchie1978one} is used to construct the electron repulsion integral, the intermediate integral can be contracted with the density matrix in the Hermite Gaussian basis. 
The Coulomb matrix is then constructed in the Hermite Gaussian basis before being transformed into the spherical harmonic Gaussian basis.
This technique, known as the "J-engine", \cite{white1996aj, shao2000improved, ahmadi1995coulomb} significantly accelerates the construction of the Coulomb matrix, primarily due to the avoidance of the quartic loop required in Electron Repulsion Integral (ERI) calculations, while maintaining exact numerical accuracy. This approach achieves $\mathcal O(N^2)$ scaling.

Seminal work by White and Head-Gordon \cite{white1994derivation, white1994continuous} on the Continuous Fast Multipole Method (CFMM) has enabled the $\mathcal O(N)$ construction of the Coulomb matrix.
The concept has been used in the development of GvFMM by Scuseria and co-workers. \cite{strain1996achieving}
In this approach, the integrals are split into "near-field" (NF) and "far-field" (FF) based on the well-separatedness criterion between interacting charge distributions.
The NF contribution scales linearly with the system size, while the FF contributions can be made linear with a hierarchical, recursive boxing algorithm. 
The NF contribution can be accelerated with the techniques discussed above.
Numerous subsequent works have incorporated the CFMM into DFT calculations for both molecular \cite{fusti2005fast, neese2025bubblepole} and periodic systems. \cite{kudin2000linear, kudin1998fast, becker2019density, lazarski2015density, lazarski2016density}

Density fitting (DF) approximation or the resolution of the identity (RI) with GTOs to approximate the ERI is another popular approach. \cite{whitten1973coulombic}
The use of RI in self-consistent field (SCF) calculation was first done by Baerends, \cite{baerends1973self} and specific improvements to RI approximation for Coulomb matrix calculation (RI-J) have been carried out in many subsequent works. \cite{franchini2014accurate, neese2003improvement, burow2009resolution, sierka2003fast, sodt2006linear}
Although asymptotically it scales as $\mathcal {O} (N^3)$, a significant speedup can be achieved for small-to-medium-sized molecules due to a smaller prefactor in the overall algorithm.
In tandem with algorithmic advancement, high-quality auxiliary basis sets have been developed by Ahlrichs, \cite{eichkorn1995auxiliary} Weigend, \cite{weigend2002fully} and other groups \cite{stoychev2017automatic} to improve the accuracy of RI-J.
In addition to the GTO-DF approximation, a technique using PWs as the DF basis (i.e., PW-DF) to efficiently evaluate the smooth density contribution to the Coulomb matrix is known as Fourier-transform Coulomb (FTC), first proposed by Pulay and coworkers. \cite{fusti-molnar_fourier_2002,fusti2002accurate, baker2004parallel} 
A similar idea was used in the Gaussian-Planewave (GPW) method popularized by CP2K developers, which readily scales as $\mathcal O(N)$.~\cite{vandevondele2005quickstep}

For extended systems, the long-range nature of the Coulomb interaction leads to the divergence of the Coulomb matrix unless the density distribution is chargeless.
The Ewald summation method, which involves calculating the long-range and short-range components independently, is commonly used to treat the divergence. \cite{delley_fast_1996, blum2009ab, patterson2020density, ye_fast_2021,sun2023exact, robinson2025} 
In practice, contributions that involve long-range interaction or smooth density are generally calculated in reciprocal space, that is, employing PWs as the fitting basis, due to their rapid convergence. \cite{lee2021approaching, sun2017gaussian}
Furthermore, the $\mathbf{G} = \mathbf{0}$ contribution, which corresponds to monopole-monopole interaction, is removed. 
The short-range interaction contribution is calculated based on analytic integration in real space, \cite{dovesi1983treatment, saunders1992electrostatic} and, for that, molecular techniques can be adapted to accelerate the computation. \cite{wang2024fast}

In this work, we present an efficient implementation of the periodic all-electron Coulomb matrix using the Ewald summation method combined with the shell classification used in FTC~\cite{fusti-molnar_fourier_2002}  and GTO-DF for the short-range contributions.
This combination was explored in the range-separated DF (RSDF) proposed by Ye and Berkelbach.~\cite{ye_fast_2021}
Compared to a na{\"i}ve use of RSDF for the Coulomb build, our approach is different in three ways, as we will elaborate more later:
\begin{enumerate}
\item For short-range integrals, we keep the shell pairs in real space using the Born–von Kármán (BvK) supercell format. 
This removes the cost associated with Fourier transforming short-range integrals from real space to reciprocal space, while enabling sparse contraction between shell pairs and the density matrix.
The real space representation has been explored in previous work. \cite{sun2023exact, lazarski2015density}
\item We also present an algorithm that leverages the product separability of shell pairs to evaluate the PW-DF of shell pairs efficiently. This makes the evaluation of the long-range component far more efficient than those that do not utilize product separability, such as RSDF. Furthermore, we do not need to store the PW-fitted intermediates.
This was motivated by the implementation of the GPW method in \texttt{CP2K}.~\cite{vandevondele2005quickstep}
\item As is done in the original FTC,~\cite{fusti-molnar_fourier_2002} the diffuse-diffuse shell pair density is only fitted by PWs, and GTO-DF basis functions are not used. Furthermore, range separation is not performed in this case. 
\end{enumerate}
The impact of these differences on the efficiency will be discussed in our timing comparison section later.

Our paper is organized as follows: (1) we review fundamentals of the periodic Coulomb matrix, (2) we present our algorithm for evaluating the Coulomb matrix, focusing on the fast evaluation of the long-range component, (3) we present numerical results, timing, and memory benchmarks on simple solids, (4) we show illustrative calculations on benzene crystal and CO adsorption on MgO (001) surface, and (5) we conclude.

\section{Theory and implementation} \label{sec:theory and implementation}

\subsection{Preliminary notations and definitions}
In periodic quantum chemistry simulations with GTOs, one works with the atomic Bloch orbitals,
\begin{equation} \label{eq:Bloch orbital}
    \phi_{\mu}^{\textbf{k}}(\mathbf{r}) = \frac{1}{\sqrt{N_k}} \sum_{\mathbf{n}} \phi_{\mu}^{\textbf{n}}(\mathbf{r}) \mathrm{e}^{i\ka \mathbf R_{\textbf{n}}},
\end{equation}
where $\mu$ indexes the $\mu$-th atomic orbital in the unit cell, 
$\phi_{\mu}^{\textbf{n}}(\mathbf{r})  = \phi_{\mu}(\mathbf{r}-\mathbf R_\textbf{n})$,
$\mathbf R_\mathbf{n}$ is the direct translation vector corresponding to unit cell $\textbf{n}$, $\ka$ is crystal momentum, and $N_k$ is the number of $\ka$ points. These symmetry-adapted orbitals satisfy Bloch theorem, $\phi_{\mu}^{\textbf{k}}(\mathbf{r} + \mathbf R_\textbf{m}) = \phi_{\mu}^{\textbf{k}}(\mathbf{r}) e^{i \ka \mathbf R_\textbf{m}}$. In a typical setting, $\ka$ is sampled uniformly from the first Brillouin zone using the Monkhorst-Pack scheme, which we shall assume throughout this paper.~\cite{monkhorst1976special} To reach the thermodynamic limit, one needs to sample more $\ka$-points from the first Brillouin zone or increase the computational unit cell size. 
During self-consistent-field (SCF) calculations, forming each contribution in $\ka$-space is typically preferred because momentum-conserving tensors become block-sparse, enabling efficient storage and diagonalization.
However, constructing $\ka$-point quantities requires performing a lattice sum (i.e., the summation over $\mathbf n$ in \cref{eq:Bloch orbital}) for real-space tensors.

For ERIs with $\ka$-point symmetry, three independent lattice summations need to be carried out
\begin{align}\nonumber
(\mu {\mathbf k}_{1} \nu {\mathbf k}_{2} | \lambda {\mathbf k}_{3} \sigma {\mathbf k}_4) =& \frac{1}{\mathrm{N}_k^{3/2}}  \sum_{\mathbf{n}} \sum_{\mathbf{m}} \sum_{\mathbf{l}} 
(\mu \mathbf{n} \nu \mathbf{m} | \lambda \mathbf{l} \sigma \mathbf{0})\\
&\times e^{i{\mathbf k}_1 \mathbf{R}_\mathbf{n}} e^{i{\mathbf k}_2 \mathbf{R}_\mathbf{m}} e^{i{\mathbf k}_3 \mathbf{R}_\mathbf{l}},
\label{eq:eri}
\end{align}
where the crystal momentum conservation demands $\ka_4 = \ka_1 - \ka_2 + \ka_3$ and $(\mu \mathbf{n} \nu \mathbf{m} | \lambda \mathbf{l} \sigma \mathbf{0})$ is defined as 
\begin{equation}
(\mu \mathbf{n} \nu \mathbf{m} | \lambda \mathbf{l} \sigma \mathbf{0})
=
\int \mathrm{d}\mathbf r_1
\int \mathrm{d}\mathbf r_2
\frac{
\phi_\mu^{\mathbf n}(\mathbf r_1)
\phi_\nu^{\mathbf m}(\mathbf r_1)
\phi_\lambda^{\mathbf l}(\mathbf r_2)
\phi_\sigma^{\mathbf 0}(\mathbf r_2)
}{|\mathbf r_1 - \mathbf r_2|}.
\end{equation}
The triple lattice summation in \cref{eq:eri} can introduce significant computational overhead for integral generation, potentially making it one of the most demanding computational steps in SCF calculations.

\subsection{Overview of Coulomb matrix computations}
In periodic SCF calculations, the Coulomb matrix, $\mathbf {J}^{\mathbf k}$, is given as 
\begin{equation} \label{eq:J matrix in Bloch orbital}
    J_{\mu\nu}^{\mathbf{k}} = \sum_{\mathbf{k'}} \sum_{\lambda \sigma} (\mu \mathbf{k} \nu \mathbf{k} | \lambda \mathbf{k'} \sigma \mathbf{k'}) \cdot D_{\lambda \sigma}^{\mathbf{k'}}.
\end{equation}
This equation suggests that calculating the Coulomb matrix directly in the Bloch orbital basis needs some care because the naive computational cost scales as $\mathcal O(N_k^2)$.
Furthermore, when using GTO-DF, the Fourier transformation of intermediate three-center two-electron integrals in real-space to the Bloch basis also scales as $\mathcal O(N_k^2)$. 
This can be seen by inspecting the following,
\begin{equation}
    (P \tilde{\mathbf 0}| \mu \mathbf k \nu \mathbf k) = \sum_{\mathbf n \mathbf m \in \text{Bvk}} (P \mathbf m| \mu \mathbf 0 \nu \mathbf n)
    e^{i\mathbf k \mathbf R_\mathbf{n}},
\label{eq:naive ft}
\end{equation}
where we used $\tilde{\mathbf 0}$ to denote $\mathbf k = \mathbf 0$ (differentiating from $\mathbf R = \mathbf 0$) and $\mathbf n$, $\mathbf m$ index BvK cells. 
Hence, we generate the Coulomb matrix in the BvK supercell, $J_{\mu\nu}^{\mathbf{0}\mathbf{n}}$, and Fourier transform it to Bloch orbital basis as follows,
\begin{equation} \label{eq:FT of J}
    J_{\mu\nu}^{\mathbf{k}} = \sum_{\mathbf{n}\in\text{BvK}} e^{i \mathbf{k} \mathbf{R}_{\mathbf{n}} } J_{\mu\nu}^{\mathbf{0}\mathbf{n}}.
\end{equation}
With the uniform $\mathbf k$-mesh,  one could use the fast Fourier transformation (FFT) for \cref{eq:FT of J}, forming an overall $\mathcal O(N_k \mathrm{log}N_k)$-scaling computation.
This ensures that even for a very dense $\mathbf k$-mesh, our Coulomb matrix formation scales at most $\tilde{\mathcal{O}}(N_k)$. 

The $\mathbf k$-space density matrix is transformed to a real-space representation via
\begin{equation} \label{eq:FT for D}
    D_{\mu\nu}^{\mathbf{0}\mathbf{n}} = \sum_{\mathbf{k}} e^{i\mathbf{k} \mathbf{R}_{\mathbf{n}}}  D_{\mu\nu}^{\mathbf{k}},
\end{equation}
where $\mathbf{n}$ are images within the BvK supercell. 
This step can also be done with FFT in each SCF iteration at the cost of $\tilde{\mathcal O}(N_k)$. 
We note that the density matrix satisfies $D_{\mu\nu}^{\mathbf{m}\mathbf{n}} = D_{\mu\nu}^{\mathbf{0}(\mathbf{n} -\mathbf{m})}$ and $D_{\mu\nu}^{\mathbf{0}\mathbf{n}} = D_{\mu\nu}^{\mathbf{0}(\mathbf{n} + \mathbf{N})}$, where $\mathbf{N}$ is the direct lattice vector of the BvK supercell, and does not decay even for large $\mathbf N$ if $\mathbf n$ is small.
This is a numerical artifact of periodic boundary conditions.
The magnitude of the density matrix should decay to zero as the boundary of the BvK supercell is approached, with an increasing number of $\mathbf k$-points being sampled. \cite{irmler2018robust}
With a relatively small size $\mathbf k$-mesh, efficient algorithms based on density matrix screening are 
hence not easily applicable.

In our implementation, the electron-electron repulsion and the electron-nuclear attraction are treated separately. 
Each of the energy contributions diverges since the corresponding density has a non-vanishing monopole. \cite{ihm1979momentum}
To ensure numerical stability, we employ the Ewald summation, also known as the range-separated Coulomb approach, and remove the diverging contribution from both terms. 
Within the Ewald summation technique, the Coulomb interaction is separated into short-range and long-range components, which are evaluated in real space and reciprocal space, respectively. 
To accelerate the evaluation of short-range integrals, we employ a \textit{global} GTO-DF with an attenuated Coulomb metric. The BvK Coulomb matrix reads
\begin{equation} \label{eq:Ewald J}
\begin{split}
    J^{\mathbf{0} \mathbf{n}}_{\mu\nu} &= \sum_{\lambda \sigma} \sum_{\mathbf{m}} \sum_{PQ} (\mu \mathbf{0} \nu \mathbf{n} | P \tilde{\mathbf{0}})_{\omega} ( P \tilde{\mathbf{0}} | Q \tilde{\mathbf{0}})_{\omega}^{-1} (Q \tilde{\mathbf{0}} | \lambda \mathbf{0} \sigma \mathbf{m} )_{\omega} \cdot D^{\mathbf{0} \mathbf{m}}_{\lambda \sigma } \\
    & + \frac{1}{\Omega} \sum_{\lambda \sigma} \sum_{\mathbf{m}} \sum_{\mathbf{G} \neq \mathbf{0}} (\mu \mathbf{0} \nu \mathbf{n}| \mathbf{G} ) \frac{4\pi}{|\mathbf{G}|^2} e^{-\frac{|\mathbf{G}|^2}{4\omega^2}} (-\mathbf{G} | \lambda \mathbf{0} \sigma \mathbf{m}) \cdot D_{\lambda \sigma }^{\mathbf{0} \mathbf{m}},
\end{split}
\end{equation}
where $\Omega$ is the unitcell volume, $\omega$ is the range-separated parameter, and $\mathbf G $ is the PW momentum vector or the reciprocal lattice vector translation. The first term amounts to the short-range contributions, and the second term accounts for the long-range contributions. The inclusion of the $\omega$ subscript denotes short-range Coulomb integrals. Details of the derivation for short-range contributions can be found in \cref{sec:J-build SR derivation}.

\subsection{Short-range contributions} \label{subsec: short-range contrib}
For the short-range component, the GTO-DF we use is robust based on the Dunlap criteria. \cite{dunlap2000robust} The short-range two-electron two-center (2e2c) integral is given by
\begin{equation} \label{eq:2e2c SR}
    (P\tilde{\mathbf{0}} | Q \tilde{\mathbf{0}})_{\omega} = \sum_{\mathbf{m}} (P \mathbf{0} | Q \mathbf{m})_{\omega} - \frac{\pi}{\Omega \omega^2} \cdot Q_P Q_Q,
\end{equation}
where $\{\mathbf{m}\}$ are significant images in real space, $\Omega$ is the volume of the unit cell, and the auxiliary shell charge is defined as
\begin{equation} \label{eq:Charge of the aux}
    Q_P = \int \mathrm{d} \mathbf{r} \phi_{P}^{\mathbf{0}}(\mathbf{r}).
\end{equation}
This charge subtraction corresponds to neglecting the $\mathbf G= \mathbf{0}$ contribution in the reciprocal space.
With the spherical harmonic Gaussian basis, this integral is $0$ unless the auxiliary shell is of s-type ($l = 0$).
The evaluation of the short-range 2e2c integrals at the $\Gamma$ point is done once at the beginning of SCF calculations and stored in memory. 

The short-range two-electron three-center (2e3c) integral is given by
\begin{equation} \label{eq:2e3c SR}
    (P \tilde{\mathbf{0}} | \mu \mathbf{0}\nu \mathbf{n})_{\omega} = \sum_{\mathbf{m}} \sum_{\mathbf{N}}
    \left((P \mathbf{m} | \mu \mathbf{0} \nu (\mathbf{n} + \mathbf{N}))_{\omega} - \frac{\pi}{\Omega \omega^2} Q_{\mu\nu}^{\mathbf{n} + \mathbf{N}} Q_{P}\right),
\end{equation}
where the charge of the pair density is defined as
\begin{equation} \label{eq:charge of 2e3c}
    Q_{\mu\nu}^{\mathbf{n} + \mathbf{N}} = \int \mathrm{d} \mathbf{r} \phi_{\mu}^{\mathbf{0}}(\mathbf{r}) \phi_{\nu}^{\mathbf{n} + \mathbf{N}}(\mathbf{r}).
\end{equation}
Here, as before, $\mathbf N$ is the direct lattice vector of the BvK supercell.

Both 2e2c and 2e3c integrals are evaluated with the Head-Gordon Pople algorithm. \cite{head1988method} The Boys function needed for s-type integral is evaluated using a modified Chebyshev table. \cite{gill1991two} 
Transformation between Cartesian and Spherical Gaussian functions is performed for batches of short-range integrals with the coefficients presented in Ref.~\citenum{schlegel1995transformation}.

For the Coulomb matrix computation, we must contract these integrals with the SCF density matrix.
An efficient contraction order for the short-range component is
\begin{enumerate}
    \item Evaluate $ W_1(Q) = \sum_{\lambda \sigma} \sum_{\mathbf{m}\in\text{BvK}} (Q\tilde{\mathbf{0}} | \lambda \mathbf{0} \sigma \mathbf{m})_{\omega} \cdot D_{\lambda \sigma}^{\mathbf{0}\mathbf{m}}$
    \item Evaluate $ W_2(P) = \text{Solve}[(P\tilde{\mathbf{0}} | Q \tilde{\mathbf{0}})_{\omega}, W_1(Q)]$
    \item Evaluate $J_{\mu\nu}^{\mathbf{0} \mathbf{n}} \mathrel{+}= \sum_{P} (\mu \mathbf{0} \nu \mathbf{n} | P \tilde{\mathbf{0}})_{\omega} \cdot W_2(P)$.
\end{enumerate}
where Solve is the linear equation solve step to effectively obtain
$W_2(P) = \sum_Q (P\tilde{\mathbf{0}} | Q \tilde{\mathbf{0}})_{\omega}^{-1} W_1(Q)$ with some
care for linear dependencies in $(P\tilde{\mathbf{0}} | Q \tilde{\mathbf{0}})_{\omega}$.
We note that solving linear equation to obtain $W_2(P)$ can be numerically unstable if $(P\tilde{\mathbf{0}} | Q \tilde{\mathbf{0}})_{\omega}$ is ill-conditioned.
Therefore, having very diffuse DF basis functions, which makes $(P\tilde{\mathbf{0}} | Q \tilde{\mathbf{0}})_{\omega}$ significantly ill-conditioned, is generally not recommended.
This is reasonable since these diffuse fitting basis functionals are likely not needed when fitting only short-range contributions.

For systems considered in this work, our timing profile indicates that performing the lattice summation in \cref{eq:2e3c SR} is typically more expensive than the subsequent contraction of 2e3c integrals with the density matrix. 
Therefore, we precalculate the 2e3c integrals (in the BvK format) at the beginning of the SCF cycle \cite{bintrim2022integral} and store them in memory with the shell pair sparsity. 
The shell pair is prescreened with a threshold of $10^{-14}$, \cite{dinh2025distance} and only significant shell pairs within a BvK cell are retained.

\subsection{Long-range contributions} \label{subsec: long-range contrib}

For the long-range component, we need to obtain the PW fitted pair density,
\begin{equation} \label{eq:AFT for pair density}
    (\mu \mathbf{0} \nu \mathbf{n}| \mathbf{G} ) = \int \mathrm{d} \mathbf{r} \rho_{\mu\nu}^{\mathbf{0}\mathbf{n}}(\mathbf{r}) e^{-i \mathbf{G} \mathbf{r}}.
\end{equation}
With the pair density, an efficient contraction order is as follows:
\begin{enumerate} \label{contraction order for AFT LR}
    \item Evaluate $\rho(-\mathbf{G}) = \sum_{\mathbf{m}\in \text{BvK}} \sum_{\lambda \sigma} (-\mathbf{G} | \lambda \mathbf{0} \sigma \mathbf{m}) \cdot D_{\lambda \sigma}^{\mathbf{0} \mathbf{m}}$
    \item Evaluate $V(-\mathbf{G}) = \rho(-\mathbf{G}) \cdot \frac{4\pi}{|\mathbf{G}|^2} \cdot e^{-\frac{|\mathbf{G}|^2}{4\omega^2}}$
    \item Evaluate $J_{\mu\nu}^{\mathbf{0}\mathbf{n}} \mathrel{+}=  \sum_{\mathbf G \ne 0}(\mu \mathbf{0} \nu \mathbf{n}| \mathbf{G} ) \cdot V(-\mathbf{G})$.
\end{enumerate}
For this part of the algorithm, it is important to perform the Fourier transformation of the pair density in \cref{eq:AFT for pair density} efficiently.
Inspired by the efficient on-the-fly evaluation of real-space density in the GPW method,~\cite{vandevondele2005quickstep} we formed an efficient contraction scheme for reciprocal-space density.

We will first review fundamental properties of GTOs and then outline the contraction algorithm.
We use $\mathbf{G} = [G_x, G_y, G_z]$ for Cartesian coordinate and $\tilde{\mathbf{G}} = [G_i, G_j, G_k]$ for relative coordinate (i.e., $\tilde{\mathbf{G}} = G_i \mathbf b_i + G_j \mathbf b_j + G_k \mathbf b_k$ with reciprocal lattice vectors, $\{\mathbf b_i\}$).
Transformation between these coordinates is discussed in \cref{sec:cart coord vs rel coord}.
We start with the primitive Cartesian Gaussian function $\phi_{a}(\mathbf{r})$ with angular momentum triplet $(l_{x_a}, l_{y_a}, l_{z_a})$, such that $l_{x_a} + l_{y_a} + l_{z_a} = l_a$,
\begin{equation} \label{eq:cart gaussian function}
    \phi_{a}(\mathbf{r}) = N_{a} \cdot x_a^{l_{x_a}} \cdot y_a^{l_{y_a}} \cdot z_{a}^{l_{z_a}} \cdot e^{-\alpha_a (x_{a}^2 + y_{a}^2 + z_{a}^2)},
\end{equation}
where $N_{a}$ is the normalization factor, and $x_{a} = x - X_{a}$, $y_{a} = y - Y_{a}$, $z_{a} = z - Z_{a}$. Here, $X_{a}, Y_{a},$ and $Z_{a}$ are nuclear center coordinates. The analytical Fourier transformation (AFT) of this primitive Gaussian function is given by
\begin{equation} \label{eq:FT of cart Gaussians}
\begin{split}
    \phi_{a}(\mathbf{G}) = &N_{a} \cdot H_{l_{x_a}} \left(\frac{G_x}{2 \sqrt{\alpha_a}}\right) H_{l_{y_a}} \left(\frac{G_y}{2 \sqrt{\alpha_a}}\right) H_{l_{z_a}} \left(\frac{G_z}{2 \sqrt{\alpha_a}}\right) \\
    & \cdot \left(\frac{\pi}{\alpha_a}\right)^{3/2} \cdot \left(\frac{i}{2\sqrt{\alpha_a}}\right)^{l_a} \cdot e^{-i(G_x X_a + G_y Y_a + G_z Z_a)} e^{-\frac{\mathbf{G}^2}{4 \alpha_a}},
\end{split}
\end{equation}
where $H_l(G)$ is the physicist Hermite polynomial of order $l$. Details of this polynomial are given in \cref{sec:hermite polynomials}.  
Using the Gaussian product theorem for a pair of primitive shells, for each Cartesian component, we have the product of the angular and radial components, 
\begin{equation} \label{eq:Gaussian product of angular prefactor}
    x_{a}^{l_{x_a}} \cdot x_{b}^{l_{x_b}} = \sum_{l_{x_p} = 0}^{l_{x_a} + l_{x_b}} x_p^{l_{x_p}} \cdot \sum_{i=0,l_{x_a}}^{i+j=l_{x_p}} \sum_{j=0, l_{x_b}}\binom{l_{x_a}}{i}\binom{l_{x_b}}{j} X_{pa}^{l_{x_a}-i} X_{pb}^{l_{x_b}-j},
\end{equation}
\begin{equation} \label{eq:Gaussian product of the radial decay}
    e^{-\alpha_{a} x_a^2} e^{-\alpha_{b} x_{b}^2} = e^{-\eta_{ab} X_{ab}^2} e^{-(\alpha_{a} + \alpha_{b}) x_p^2},
\end{equation}
where
\begin{equation} \label{eq:definition of GPT}
    \eta_{ab} = \left[ \frac{1}{\alpha_{a}} + \frac{1}{\alpha_{b}} \right]^{-1},
\end{equation}
and the location of the new Gaussian center is
\begin{equation}
    X_p = \frac{\alpha_{a} X_{a} + \alpha_{b} X_{b}}{\alpha_{a} + \alpha_{b}}.
\end{equation}
The Gaussian function product is another Gaussian function, and we can apply \cref{eq:FT of cart Gaussians} to obtain its Fourier transformation. 

Since the PW fitted pair density does not need to be explicitly calculated in the Coulomb matrix construction, we perform the contraction for integral intermediates on the fly.
One crucial observation is that the computational cost can be minimized if we work in the relative frame of the reciprocal space lattice vector. 
The derivation of the full contraction implemented in this work can be found in \cref{sec:eval of density on reciprocal space}.

Given these, an efficient on-the-fly contraction for building $\rho(\mathbf{G})$ is presented in \cref{alg: non-ortho rho eval}. 
The evaluation of the periodic Coulomb matrix from the potential $V(\mathbf{G})$ follows a similar loop structure and is presented in \cref{sec:J build acceleration for grid based}. 
In each primitive shellpair loop, the intermediate along each relative coordinate,
\begin{equation} \label{eq:intermediate in relative coordinate}
    I_{l_{G_i}}(G_i) = G_i^{l_{G_i}} \cdot e^{-i 2\pi G_i I_p},
\end{equation}
is computed recursively using the relation $I_{l}(G_i) = G_i \cdot I_{l-1}(G_i)$. We have $l_{G_i}=\left\{0,1, \ldots, l_a^{\max }+l_b^{\max }\right\}$ within a shell, and $I_p$ is the relative coordinate of the primitive shell pair center.
The partially separable exponential intermediate,
\begin{equation} \label{eq:non-separable exponential term}
    E(G_i, G_j) = \exp \left[-\frac{1}{4\alpha_p} \left( \mathbf{b}_i \cdot \mathbf{b}_i G_i^2 + 2 \mathbf{b}_i \cdot \mathbf{b}_j G_i G_j \right) \right],
\end{equation}
is evaluated on a two-dimensional PW basis. 
To avoid expensive exponential function calls on a two-dimensional grid, we construct $E(G_i, G_j)$ as follows.
Consider the term $\text{exp}(- G_i G_j)$ (ignoring other constant factors on the exponent for simplicity), we first calculate the 
$\text{exp}(-G_i G_j^{\text{min}})$ and $\text{exp}(- G_i \delta G_j)$, called the shift term, where $G_j^{\text{min}}$ is the minimum coordinate along vector $\mathbf{b}_j$ and $\delta G_j$ is the spacing between uniformly sampled PWs along $\mathbf{b}_j$.
Given a fixed $G_i$, we loop through $G_j$ grid points to incrementally update $\text{exp}(-G_i G_j)$ with the shift term, starting from $G_j^{\text{min}}$.
Therefore, the total number of exponential calls per primitive shell pair is $\mathcal{O}(N_{G_i} + N_{G_j} + N_{G_k}) = \mathcal{O}(N_G^{1/3})$, where $N_{G} = N_{G_i} N_{G_j} N_{G_k}$ is the total number of PWs.

\begin{algorithm} [H]
\caption{Evaluate $\rho(\mathbf{G})$ given $D_{\mu\nu}^{\mathbf{0}\mathbf{n}}$} \label{alg: non-ortho rho eval}
\begin{algorithmic}
    \FOR{primitive shell pair in $| \mu \mathbf{0} \nu \mathbf{n})$}
        \STATE Evaluate $I_{l_{G_i}}(\mathbf{G}_i)$, $I_{l_{G_j}}(\mathbf{G}_j)$, $I_{l_{G_k}}(\mathbf{G}_k)$.
        \STATE Evaluate $E(\mathbf{G}_i, \mathbf{G}_j)$, $E(\mathbf{G}_j, \mathbf{G}_k)$, $E(\mathbf{G}_k, \mathbf{G}_i)$.
        \STATE Use \cref{alg: density matrix transformation} to obtain $D_{l_{G_i}, l_{G_j}, l_{G_k}}$.
        \FOR{$G_k$ in $\{\mathbf{G}_k\}$}
            \STATE $D_{l_{G_i}, l_{G_j}}(G_k) \leftarrow \sum_{l_{G_k}} D_{l_{G_i}, l_{G_j}, l_{G_k}} \cdot I_{l_{G_k}}(G_k)$
            \FOR{$G_j$ in $\{\mathbf{G}_j\}$}
                \STATE $I_{l_{G_j}}(G_j, G_k) \leftarrow I_{l_{G_j}}(G_j) \cdot E(G_j, G_k)$
                \STATE $D_{l_{G_i}}(G_j, G_k) \leftarrow \sum_{l_{G_j}} D_{l_{G_i}, l_{G_j}}(G_k) \cdot I_{l_{G_j}}(G_j, G_k)$
                \FOR{$G_i$ in $\{\mathbf{G}_i\}$}
                    \STATE $I_{l_{G_i}}(G_i, G_j, G_k) \leftarrow I_{l_{G_i}}(G_i) \cdot E(G_i, G_j) \cdot E(G_k, G_i)$
                    \STATE $\rho(G_i, G_j, G_k) \mathrel{+}= \sum_{l_{G_i}} D_{l_{G_i}}(G_j, G_k) \cdot I_{l_{G_i}}(G_i, G_j, G_k)$             
                \ENDFOR
            \ENDFOR
        \ENDFOR
    \ENDFOR
\end{algorithmic}
\end{algorithm}

The bottleneck comes from the three nested loops over each PW direction. 
The asymptotic scaling of the procedure presented here is $\mathcal{O}((l_{a}^{\max} + l_{b}^{\max} + 1) N_{\text{prim}} N_{G})$, where $N_{\text{prim}}$ is the number of primitive shell pairs. 
Even though the asymptotic scaling does not change from the naive contraction between $\rho_{\mu\nu}^{\mathbf{0}\mathbf{n}}(\mathbf{G})$ and $D_{\mu\nu}^{\mathbf{0}\mathbf{n}}$.
\cref{alg: non-ortho rho eval} is substantially faster due to (1) reduced prefactor from $(2l_a + 1) (2l_b + 1)$ to $l_a + l_b +1$, (2) the avoidance of performing explicit exponential evaluations at each PW for forming $\rho_{\mu\nu}^{\mathbf{0}\mathbf{n}}(\mathbf{G})$, and (3) the removal of assembling 3D Cartesian Gaussian integrals from each 1D component. 
In our internal benchmark, assembling the 1D integral component across the entire PW basis is the primary bottleneck when building PW-DF explicitly for each shell pair, which we now avoid entirely.
Furthermore, the algorithm presented here achieves even better performance when working with the multishell structure, where different atomic shells share the same set of primitives.
This is because primitive shells with different angular momentum can be processed simultaneously with the density matrix transformation given in \cref{alg: density matrix transformation}.
This multi-shell design is prevalent in many available basis sets, such as atomic natural orbital, \cite{almlof1987general} correlation-consistent basis sets, \cite{peterson2002accurate}
and the \texttt{CP2K} basis sets.~\cite{vandevondele2007gaussian}

\begin{algorithm} [H]
\caption{BvK density matrix $D_{\mu\nu}^{\mathbf{0}\mathbf{n}}$ transformation} \label{alg: density matrix transformation}
\begin{algorithmic}
    \FOR{primitive shell pair in $| \mu \mathbf{0} \nu \mathbf{n})$}
        \STATE Absorb the contraction coefficient into $D_{\mu\nu}^{\mathbf{0}\mathbf{n}}$.
        \STATE Transform $D_{\mu\nu}^{\mathbf{0}\mathbf{n}}$ to Cartesian Gaussian  basis and unfold to $D_{l_{x_a}, l_{y_a}, l_{z_a}, l_{x_b}, l_{y_b}, l_{z_b}}$.
        \STATE Transform $D_{l_{x_a}, l_{y_a}, l_{z_a}, l_{x_b}, l_{y_b}, l_{z_b}}$ with ~\cref{eq:Gaussian product of angular prefactor} to $D_{l_{x_p}, l_{y_p}, l_{z_p}}$.
        \STATE Transform $D_{l_{x_p}, l_{y_p}, l_{z_p}}$ with ~\cref{eq:FT of cart Gaussians} to $D_{l_{G_x}, l_{G_y}, l_{G_z}}$.
        \STATE Transform $D_{l_{G_x}, l_{G_y}, l_{G_z}}$ with ~\cref{eq:transform cart angular factor to relative angular factor} to $D_{l_{G_i}, l_{G_j}, l_{G_k}}$.
    \ENDFOR
\end{algorithmic}
\end{algorithm}

We now focus on the transformation of the BvK density matrix $D_{\mu\nu}^{\mathbf{0} \mathbf{n}}$ (\cref{eq:FT for D}).
As shown in \cref{alg: density matrix transformation}, this density matrix is first converted to a Cartesian Gaussian basis and then unfolded into independent 1D angular momentum components, where $l_{x_a}, l_{y_a}, l_{z_a} \in \{0, \ldots, l_a^{\max}\}$ and $l_{x_b}, l_{y_b}, l_{z_b} \in \{0, \ldots, l_b^{\max}\}$.
All the prefactors that are intrinsic to the transformation of the polynomials (\cref{eq:Gaussian product of angular prefactor}, \cref{eq:FT of cart Gaussians}, and \cref{eq:transform cart angular factor to relative angular factor}) are then employed to transform the unfolded density matrix $D_{l_{x_a}, l_{y_a}, l_{z_a}, l_{x_b}, l_{y_b}, l_{z_b}}$.
In the end, we obtain the transformed density matrix $D_{l_{G_i}, l_{G_j}, l_{G_k}}$, where $l_{G_i}, l_{G_j}, l_{G_k} \in \left\{0, \ldots, l_a^{\max }+l_b^{\max }\right\}$.
The cost of these transformations is independent of the number of grid points $N_G$.
Therefore, this part will not be the computational bottleneck.

If the unit cell is orthorhombic, where the primitive lattice vector satisfies \cref{eq:ortho condition}, $\tilde{\mathbf{G}}$ is the same as $\mathbf{G}$. 
We use recurrence relations \cite{sun2017gaussian} directly on the pair density $\rho_{\mu\nu}(\mathbf{G})$, where these intermediates are calculated for unique $G_x, G_y$, and $G_z$ only.
Due to the complete separation of different directions, one could achieve further speedup, but we do not explore a specialized production-level implementation for orthorhombic cells in this work.

\subsection{Integral screening} \label{subsec: integral screening}

We discuss the integral screening for long-range integral evaluation as described in~\cref{subsec: long-range contrib}. Efficient integral screening for short-range two-electron integrals has been reported elsewhere \cite{dinh2025distance, ye2021tight, izmaylov2006efficient} and will not be discussed here. 
In the precomputation step, for a unique contracted shell pair $[\mu\nu]$, we employ the most compact primitive pair (ab) and perform a binary search to identify the reciprocal space cutoff $G_{\text{cut}}$ for the entire shell via the error function
\begin{equation} \label{eq:AFT screening obj}
    \epsilon_{ab}(G) = 8 \pi \eta_{ab\omega} G \cdot \frac{4 \pi}{G^2} e^{-\frac{G^2}{4\omega^2}} \cdot |\rho^{\text{exact}}_{ab}(G)|,
\end{equation}
where $|\rho^{\text{exact}}_{\mu\nu}(G)|$ is the infinity norm of the exact pair density AFT and 
\begin{equation} \label{eq:eta_munu}
    \eta_{ab\omega} = \left[ \frac{1}{\alpha_{a} + \alpha_{b}} + \frac{1}{\omega^2} \right]^{-1}.
\end{equation}
The prefactor, $8 \pi \eta_{ab\omega} G$, in the above expression comes from the integral test to account for contributions outside of the bounding sphere.  
We use the exact analytical form for $\rho_{ab}(G)$ in the above screening since it is inexpensive to obtain analytical values for unique shell pairs. 
The cutoff value $G_{\text{cut}}$ is tabulated for each bin of the distance cutoff between shells in the shell pair.~\cite{ye2021tight}
We use $G_{\text{cut}}$ to determine the range of PWs as described below.

For a given offset in the reciprocal space $\mathbf{G}_{0}$, we want to find the range of the PWs along the reciprocal lattice vector $\mathbf{b}_i$ direction such that 
\begin{equation} \label{eq:G-space inequality}
    |\mathbf{G}_0 + n_i \mathbf{b}_i| \leq G_{\text{cut}},
\end{equation}
where $n_i$ is an integer. 
We obtain the upper bound $n_i^{\text{max}}$ and $n_i^{\text{min}}$ by finding the roots of the following quadratic equation
\begin{equation} \label{eq:G-space root finding}
    n_i^2 |\mathbf{b}_i|^2 + 2 n_i |\mathbf{b}_i \cdot \mathbf{G}_i| + |\mathbf{G}_0|^2 = G_{\text{cut}}^2,
\end{equation}
As long as the current offset vector $\mathbf{G}_0$ is inside the bounding sphere with radius $G_{\text{cut}}$, we have the following property $n_i^{\text{max}} \geq 0 \geq n_i^{\text{min}}$. 
Furthermore, we are guaranteed to find the range of PWs along the reciprocal lattice vector $\mathbf{b}_i$.
This procedure can be applied to find the maximal PWs range along each primitive reciprocal lattice vector direction by setting $\mathbf{G}_0 = \mathbf{0}$. 
Thus, we only evaluate the expensive exponential in the radial terms within the significant range for a primitive shell pair.

\subsection{Density classification: compact and diffuse} \label{subsec: compact diffuse}
\begin{figure}[H]
    \centering
    \includegraphics[width=\linewidth]{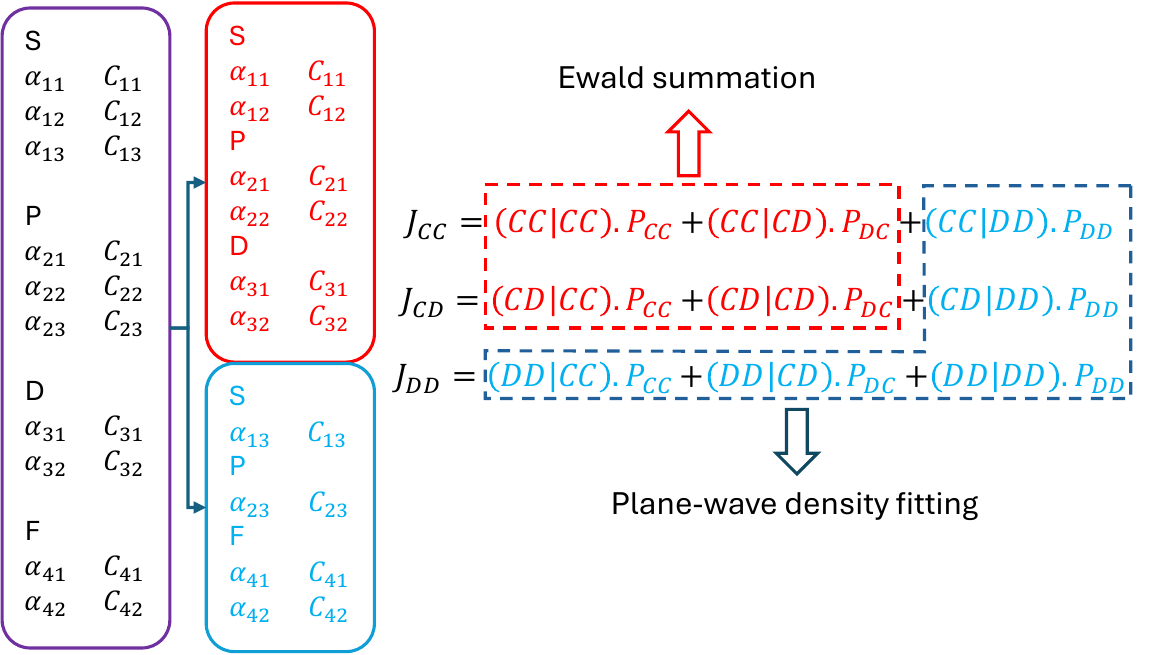}
    \caption{Workflow illustration of periodic Coulomb matrix construction. Input shells are classified as compact (C) or diffuse (D) based on the smallest exponent and are uncontracted when necessary. The Coulomb matrix is then evaluated either via Ewald summation with GTO-based DF or through pure PW-DF. On the left, we schematically show how a contracted GTO is split into compact and diffuse shells. $\alpha_{ij}$ denotes an exponent of a primitive GTO and $C_{ij}$ is the corresponding contraction coefficient.}
    \label{fig:workflow}
\end{figure}

In the Ewald summation, the cost of the 2e3c (\cref{eq:2e3c SR}) evaluation is highly dependent on the diffuseness of the basis functions and the range-separated parameter.
To reduce the workload in generating GTO-DF two-electron integrals, we classify input shells into either compact (C) or diffuse (D) based on their most diffuse exponent compared to a pre-defined cutoff exponent. \cite{fusti-molnar_fourier_2002, ye_fast_2021} 
If a contracted shell contains both compact and diffuse primitives, we uncontract the corresponding shell and split it into two separate shells to handle them individually.

In \cref{fig:workflow}, we summarize how different blocks of the Coulomb matrix are generated.  
In essence, when both sides of the integrals consist of compact pair densities (CC/CD), we employ the Ewald summation with short- and long-range integrals evaluated as in \cref{subsec: short-range contrib} and \cref{subsec: long-range contrib}. 
For all other cases, the Coulomb matrix is evaluated entirely in the reciprocal space, and the divergence term is removed through $\mathbf{G} = \mathbf{0}$ contribution. 
The Fourier transformation of the DD densities is done using the GPW \cite{lippert1997hybrid,vandevondele2005quickstep, kuhne2020cp2k} method. 
In \cref{sec:accelerate rho build GPW}, we briefly outline the contraction scheme used in the GPW approach to expedite the evaluation of the density on a uniform real space grid.
In the current implementation, we enforce a similar uniform grid size in the long-range component within the Ewald summation and in GPW.
For computational efficiency, all Coulomb potentials fitted with the PW basis, which include the long-range contribution for the compact density and the full-range contribution for the diffuse density, are combined before being passed to the Coulomb construction kernel.

When the density classification is enabled, the default cutoff exponent for compact/diffuse Gaussian shells is given as 
\begin{equation} \label{eq:cd cutoff}
    \alpha_{\text{cut}} = \frac{\omega^2}{2},
\end{equation}
where $\omega$ is the range-separated parameter. 
Compact basis functions have exponents larger than $\alpha_{\text{cut}}$, whereas diffuse basis functions have smaller exponents than $\alpha_{\text{cut}}$.
This cutoff ensures that the exponential decay of the PW-fitted diffuse density (DD block), $\exp(-|\mathbf{G}|^2/ 4(2\alpha_{\text{cut}}))$ is asymptotically similar to that of the long-range kernel, $\exp(-|\mathbf{G}|^2/ 4 \omega^2)$.
Therefore, integral classes calculated through the PW-DF converge numerically when using similar planewave densities for AFT in the Ewald summation and GPW. 

\subsection{Computational scaling for core computational kernels} \label{subsec: computational scaling}

Each computational component presented in 
\cref{subsec: short-range contrib}, \cref{subsec: long-range contrib}, and \cref{subsec: compact diffuse} have their own scaling with respect to the number of atoms ($N$) in the unit cell.
To facilitate our discussion on the scaling with respect to the supercell size, we outline the ideal scaling and current implementation scaling as a function of $N$ for core computational steps in \cref{tab: expected computational scaling}.
Within each SCF iteration, we only present the scaling for operations that involve contraction with the density matrix.
The construction of the Coulomb matrix from either PW fitted potential $V(\mathbf{G})$ or $W_2(P)$ exhibits similar scaling and therefore is omitted for simplicity.
All operations here are expected to be independent of $N_k$.

\renewcommand{\arraystretch}{1.4}
\begin{table}[H]
\captionsetup{type=table}
\captionof{table}{Expected computational scaling with respect to $N$ for various components of the algorithm.
The first three operations are included in the initialization time.
The remaining operations are carried out within each SCF iteration.}
\begin{tabular}{ C{0.38\linewidth} | C{0.28\linewidth} |  C{0.28\linewidth} }
\hline
\hline
 Operation  & \makecell[c]{Ideal \\ Scaling} & \makecell[c]{Current \\ Implementation} \\
\hline
    Obtain $(P \tilde{\mathbf{0}}| Q \tilde{\mathbf{0}})_{\omega}$ \text{and store}  &  $\mathcal{O}(N)$ & $\mathcal{O}(N^2)$  \\
    Obtain $(P \tilde{\mathbf{0}} | \mu \mathbf{0} \nu \mathbf{n})_{\omega}$ \text{and store}  &  $\mathcal{O}(N)$ & $\mathcal{O}(N^2)$ \\
\hline
    $(P \tilde{\mathbf{0}} | \mu \mathbf{0} \nu \mathbf{n})_{\omega} \cdot D_{\mu\nu}^{\mathbf{0} \mathbf{n}}$ & $\mathcal{O}(N)$ & $\mathcal{O}(N^2)$ \\
    
    \text{Solve}$[(P \tilde{\mathbf{0}} | Q \tilde{\mathbf{0}})_{\omega}, W_1(Q)]$ & $\mathcal{O}(N^3)$ & $\mathcal{O}(N^3)$ \\
    
    $\rho_{\mu\nu}^{\mathbf{0}\mathbf{n}}(\mathbf{G}) \cdot D_{\mu\nu}^{\mathbf{0} \mathbf{n}}$ & $\mathcal{O}(N^2)$ & $\mathcal{O}(N^2)$ \\
    
    $\rho_{\mu\nu}^{\mathbf{m}\mathbf{n}}(\mathbf{r}) \cdot D_{\mu\nu}^{\mathbf{m} \mathbf{n}}$ & $\mathcal{O}(N)$ & $\mathcal{O}(N)$ \\
\hline
\hline
\end{tabular}
\label{tab: expected computational scaling}
\end{table}

Ideally, the generation of the short-range two-electron integrals in \cref{eq:2e2c SR} and \cref{eq:2e3c SR} scales linearly in $N$ due to shell pair sparsity and the exponential decay of short-range interaction.
However, we do not account for the sparsity between bra and ket distributions when storing 2e2c and 2e3c integrals and for the removal of the $\mathbf{G} = \mathbf{0}$ component from these integrals (\cref{eq:2e2c SR} and \cref{eq:2e3c SR}).
Therefore, the current initialization step has asymptotic scaling of $\mathcal{O}(N^2)$, assuming the shell pair sparsity within the central unit cell.
Furthermore, the contraction of the 2e3c short-range integrals with the density matrix within each SCF iteration in our implementation also scales as $\mathcal{O}(N^2)$ instead of the ideal $\mathcal{O}(N)$.

The long-range AFT density matrix contraction has $\mathcal{O}(N^2)$ scaling. 
This quadratic scaling of the AFT component arises from the linear increase in the number of shell pairs, $N_{\text{sp}}$, and the number of PWs, $N_{\text{PW}}$. Note that for a fixed $\omega$, $N_{\text{PW}}$ increases linearly with the cell volume, which is proportional to the number of atoms.
We pick $N_{\text{PW}}$ to converge the reciprocal lattice summation as discussed in \cref{subsec: integral screening}.
Furthermore, in our default setting, we solve the linear equation to complete the GTO-DF each SCF cycle, which scales as $\mathcal{O}(N^3)$.
Note that one could take a pseudoinverse and store it in the initialization and then reuse it for SCF.
The last operation, $\rho_{\mu\nu}^{\mathbf{m}\mathbf{n}}(\mathbf{r}) \cdot D_{\mu\nu}^{\mathbf{m} \mathbf{n}}$, is used exclusively for diffuse densities with GPW method and exhibits linear scaling due to the exponential decay of Gaussian functions.
All PW-DF kernels are implemented at the production level with ideal scaling behavior.

\section{Numerical results and Discussion} \label{sec:numerical section}

The algorithm above has been implemented within a development version of \texttt{Q-Chem}, including its periodic boundary condition program, \texttt{QCPBC}. \cite{epifanovsky2021software, lee2021approaching, lee2022faster,rettig2023even} 
We also used \texttt{PySCF} \cite{sun2020recent} to validate our calculations and to compare timings with our method.
For all calculations, we used the $\text{SG}1$ grid \cite{gill1993standard} with PBE functional \cite{perdew1996generalized} unless specified otherwise.
The partition function for numerical integration of the exchange-correlation functional over the Becke atom-centered grid~\cite{becke1988multicenter} is presented in \cref{sec:XC details}.
The shell pair, 2e3c bra/ket distribution distance, 2e2c bra/ket distribution distance, and PW-DF screening thresholds are set to $10^{-14}$, $10^{-10}$, $10^{-18}$, and $10^{-12}$, respectively.
This default screening setting is used in all timing benchmarks and illustrative calculations unless specified otherwise.
All timing results were obtained using 32 threads on an AMD EPYC 9654 96-Core Processor (2.4 GHz).
Furthermore, only the non-orthorhombic implementation of the PW-DF was used for the timing/scaling benchmark.

\subsection{Numerical accuracy as a function of $\omega$}
We consider a silicon crystal for numerical accuracy tests.
We use the correlation-consistent basis set cc-pVXZ (X = D, T, and Q), \cite{dunning1989gaussian} and the auxiliary basis set is cc-pVXZ-jkfit.  \cite{weigend2002fully} 
Since the optimized auxiliary basis set for the cc-pVDZ basis set was not reported, we use cc-pVTZ-jkfit as the DF basis for this purpose. 

\begin{figure}[H]
    \centering
    \includegraphics[width=0.85\linewidth]{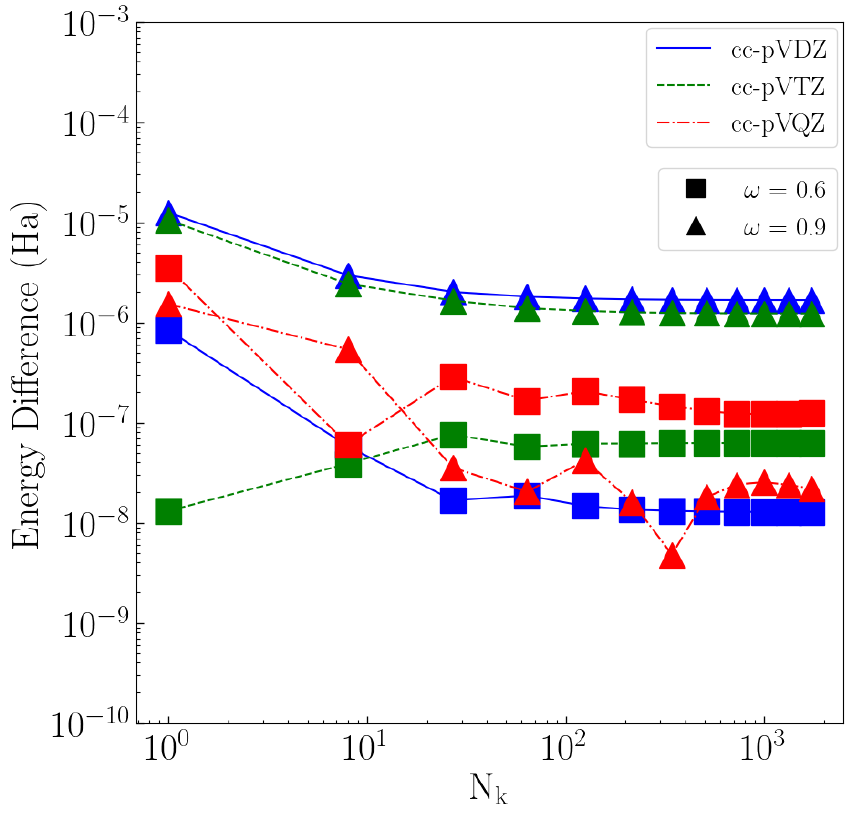}
    \caption{Difference in the converged SCF energies for the Si crystal with cc-pVDZ, cc-pVTZ, and cc-pVQZ basis sets at various range-separated parameter $\omega$ values. The energy difference is reported as the absolute difference to the SCF energies obtained with $\omega = 0.3$. The number of sampled $\mathbf k$-mesh is from $1\times1\times1$ to $12\times12\times12$.}
    \label{fig:si_crystal_energy}
\end{figure}

Our Coulomb algorithm (\cref{sec:theory and implementation}) employs both PWs and GTOs as fitting bases. When a sufficiently large number of PWs is included, the corresponding integral class is free of DF error. In contrast, the use of GTOs for DF introduces an error, which can be systematically reduced by adopting a more complete fitting basis. 
As a result, the DF error in our approach only occurs in the evaluation of short-range integrals associated with compact shell-pair densities.
A varying range-separated parameter $\omega$ leads to a different fitting metric $(P \tilde{\mathbf{0}} | Q \tilde{\mathbf{0}})_{\omega}$, which affects the accuracy of DF.
On the other hand, the cutoff exponent $\alpha_{\text{cut}}$ controls the number of pair densities that either GTOs or PWs will fit.
Therefore, converged SCF energy is dependent on $\omega$ and the cutoff exponent $\alpha_{\text{cut}}$.
The final energy would be insensitive to these parameters if the short-range integrals do not use DF, for instance, as in the J-engine technique.~\cite{white1996aj}

In ~\cref{fig:si_crystal_energy}, we observe that the energy changes for using different values of $\omega$ are less than 20 $\mu\text{Ha}$ for Si with various cc-pVXZ basis sets.
The energy difference tends to decrease as the $\mathbf k$-mesh becomes denser, indicating that the GTO-DF error is smaller as one approaches the thermodynamic limit.  
This is similar to the reduction of basis set incompleteness errors as the system approaches the thermodynamic limit, as observed in our previous study.~\cite{lee2021approaching}
Overall, the energy difference across various calculation setups is within the typical GTO-DF error (50-60 $\mu\text{Ha}$/atom).
These small changes in energy resulting from a different number of compact/diffuse basis functions also indicate the high quality of the GTO-DF basis.
We also observe similar behavior for LiF with def2 basis sets, as shown in \cref{sec:LiF SCF energies}.

\begin{figure*}
    \centering
    \includegraphics[width=\textwidth]{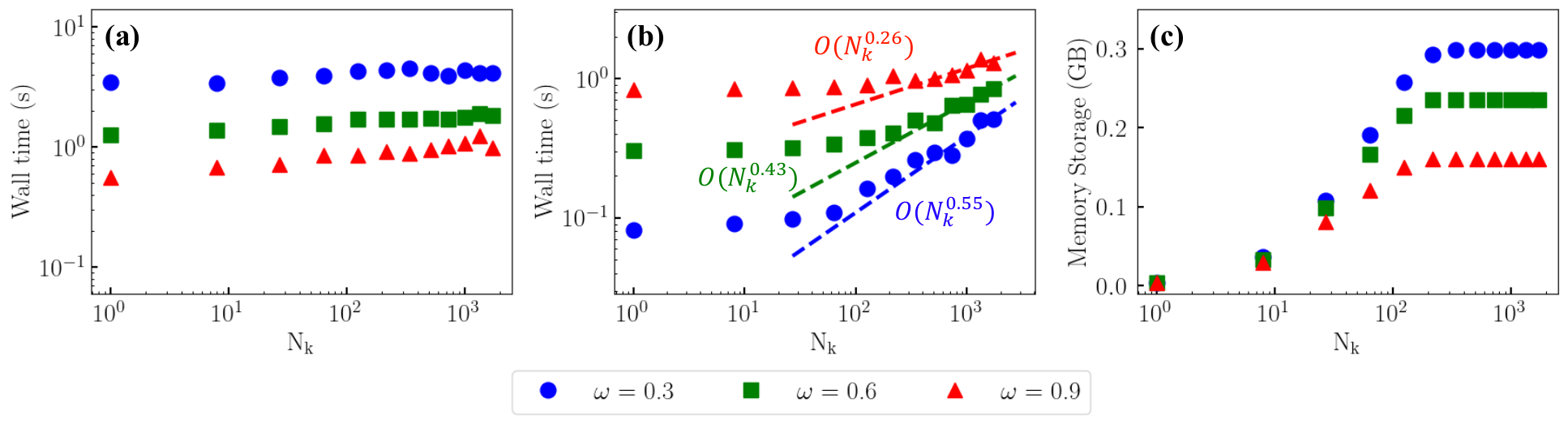}
    \caption{ \textbf{(a)} Initialization time, \textbf{(b)} time per iteration, \textbf{(c)} and 2e3c-integral storage requirement as a function of $N_k$ for Si crystal with cc-pVTZ basis set. 
    The dependence of the time per iteration on $N_k$ is obtained by fitting the wall time with the five largest $\mathbf k$-point meshes.
    The $\mathbf k$-mesh is sampled from $1\times1\times1$ to $12 \times 12 \times 12$.}
    \label{fig:si_ccpvtz_nk_composite}
\end{figure*}

\begin{figure*}
    \centering
    \includegraphics[width=\textwidth]{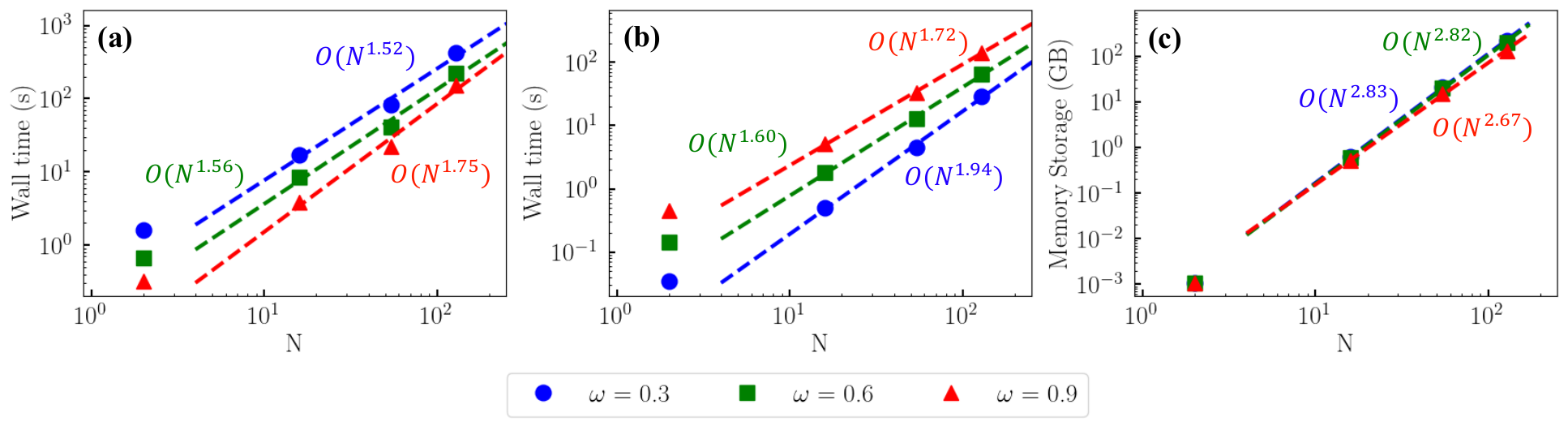}
    \caption{ \textbf{(a)} Initialization time, \textbf{(b)} time per iteration, \textbf{(c)} and 2e3c integral storage requirement as a function of the number of atoms ($N$) in the unit cell for Si crystal with cc-pVDZ basis set. 
    The dependence on $N$ is fitted using data points from (2,2,2) to (4,4,4) supercell. 
    The unit cell size is sampled from (1,1,1) to (4,4,4).
    }
    \label{fig:si_ccpvdz_supercell_composite}
\end{figure*}

\subsection{Computational scaling}

Analyzing the computational scaling of the algorithm presented here is somewhat complicated since different computational kernels are used in the calculation for each integral class. 
Instead of examining the individual scaling for each kernel, we will focus our analysis on the wall time for initialization and for J-build as a function of the range-separated parameter $\omega$, the number of $\mathbf k$-points $N_k$, and the supercell sizes $N$.
For the scaling with respect to $N_k$, we used Si with the cc-pVTZ basis set as an example.
For the scaling with respect to the supercell size, we used Si with the cc-pVDZ basis set as the test.

In ~\cref{fig:si_ccpvtz_nk_composite}(a), we report the initialization time, which consists of generating the three-center and two-center integrals, as a function of $N_k$. 
During the initialization step, we also perform an FFT of the overlap matrix from the Bloch basis to the BvK representation. Nonetheless, the cost of this step is insignificant compared to generating the two-electron integrals.
We observe that the initialization time decreases with increasing $\omega$ due to a smaller distance cutoff between bra and ket charge distributions.
Additionally, if we enable compact/diffuse classification, fewer expensive diffuse pair densities are passed through the short-range kernel, further reducing the workload.
The 2e3c integrals are generated and stored in the sparse BvK format (\cref{eq:2e3c SR}), resulting in a wall time largely independent of $N_k$.
Overall, this sparsity-aware implementation enables efficient SCF calculations on a dense $\mathbf k$-mesh, avoiding the computational overhead of performing the Fourier transformation on intermediate integrals.

In \cref{fig:si_ccpvtz_nk_composite}(b), time per J-build remains nearly constant in $N_k$ until the $\mathbf k$-mesh becomes relatively dense.
This is because beyond the $N_k$-independent operations presented in ~\cref{subsec: computational scaling}, several steps in our Coulomb implementation scale as $\mathcal{O}(N_k)$.
These include FFT of the density matrix/BvK Coulomb matrix (\cref{eq:FT for D,eq:FT of J}), and the mapping forward and backward between compact/diffuse submatrices and the whole matrix within each $\mathbf k$-point block.
In the regime where the time per iteration is $N_k$-dependent, we empirically show that the wall time scales as $\mathcal{O}(N_k^{n})$, where $n < 1$.
As $\omega$ increases, more $N_{\text{PW}}$ is needed for PW-DF, 
which further postpones the onset of the wall time growth with respect to $N_k$.
Overall, the timing presented here indicates that time per J-build scales sublinearly with $N_k$ over a broad range of $N_k$ up to 1,728.
We observe a similar $N_k$-dependence for Si with cc-pVDZ and cc-pVTZ basis sets in \cref{sec:extra data for Si}.
We note that other components of the SCF routine, such as Fock matrix diagonalization, scale formally as $\mathcal{O}(N_k)$.
Hence, the $N_k$-dependence of time per iteration presented here is unlikely to improve in the future.

In \cref{fig:si_ccpvdz_supercell_composite}(a), we show the scaling of the initialization time with the number of atoms in the unit cell, up to 128 atoms. 
The empirical scaling between linear and quadratic scaling follows our expectation in \cref{tab: expected computational scaling}.
For larger $\omega$, the lattice sum time to generate $(P \tilde{\mathbf{0}} | \mu \mathbf{0} \nu \mathbf{n})_{\omega}$, which scales as $\mathcal{O}(N)$, contributes less to the total initialization time, leading to a higher scaling because other $\mathcal O(N^2)$ steps dominate. 
This is due to our current implementation limitations (see \cref{subsec: computational scaling}).
We note that the total initialization time should eventually scale as $\mathcal{O}(N^2)$ (ideally $\mathcal{O}(N)$) when shell pair sparsity is in effect. 
However, for relatively small system sizes, the number of significant shell pairs still grows quadratically. Hence, the total initialization time can range from quadratic to cubic scaling before we reach the asymptotic regime.
In \cref{fig:si_ccpvdz_supercell_composite}(b), we observe subquadratic scaling with respect to the number of atoms,
indicating that the on-the-fly PW-DF is the dominant computational cost.
As $\omega$ increases, we classify more GTOs as diffuse functions (see \cref{eq:cd cutoff}) and hence put more contributions done by the efficient GPW code (i.e., $\mathcal O(N)$ scaling), leading to a lower scaling.

We examine the 2e3c integral storage requirement as a function of $N_k$ and $N$ in \cref{fig:si_ccpvtz_nk_composite}(c) and \cref{fig:si_ccpvdz_supercell_composite}(c).
The storage formally scales as $\mathcal{O}(N_{sp} \cdot N_{aux})$.
We observe that the memory requirements remain modest for the simple solids with dense $\mathbf k$-point sampling.
With dense $\mathbf k$-point sampling, i.e., in the limit of a large BvK supercell, the number of significant shell pairs within the BvK supercell ceases to grow, leading to a plateau in memory requirement as seen in \cref{fig:si_ccpvtz_nk_composite}(c).
The decrease in integral storage as $\omega$ increases comes from the compact/diffuse classification.
More shells are classified as diffuse when $\omega$ increases (\cref{eq:cd cutoff}), removing the storage requirement for 2e3c integrals associated with the diffuse density (DD block).  
As shown in \cref{fig:si_ccpvdz_supercell_composite}(c), the storage requirement has a steep scaling with $N$ for supercell calculation.
This is because both the auxiliary basis dimension and the primary basis dimension scale linearly with the supercell size, and we are not accounting for the sparsity between bra and ket distribution when storing the integral.
The quadratic to cubic storage indicates that the shell pair sparsity has not reached the linear scaling regime.
In the future, one could implement an on-the-fly construction of the short-range Coulomb matrix (i.e., integral-direct) to remove the memory bottleneck associated with supercell simulations.

\begin{figure}[H]
    \centering
    \includegraphics[width=0.8\linewidth]{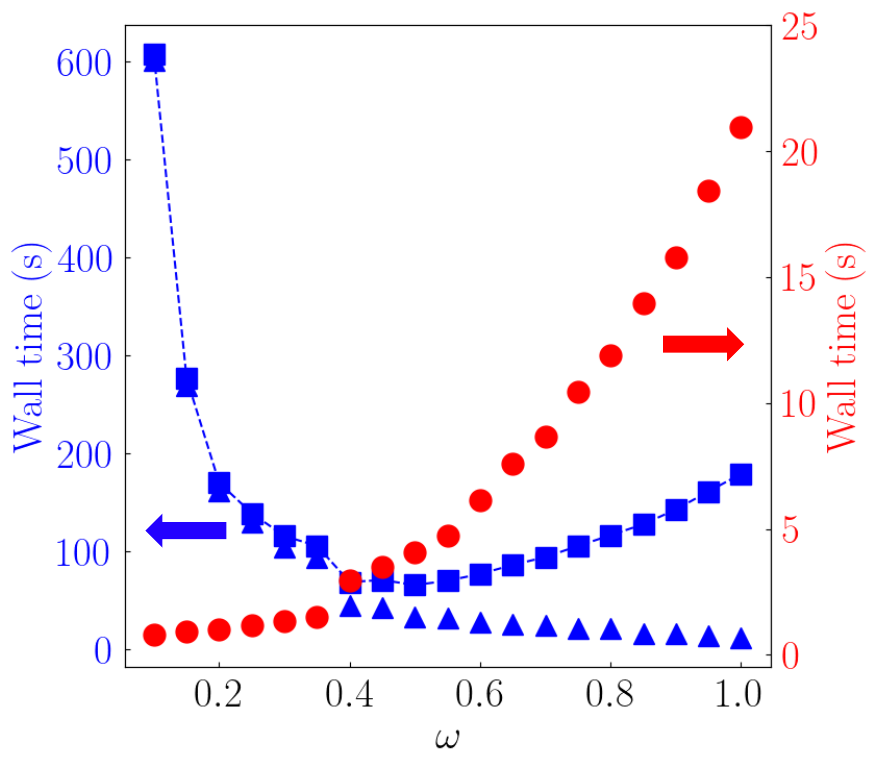}
    \caption{Dependence of SCF wall time on the range-separation parameter $\omega$
    for a Si (2,2,2) supercell with the cc-pVTZ basis set. The two-electron integral generation time (left) and per-iteration time (right) are shown. The total RI-J time, which consists of 8 iterations, is indicated by square markers. Values of $\omega$ are sampled from 0.1 to 1.0.}
    \label{fig:Si ccpvtz omega dependence}
\end{figure}

In ~\cref{fig:Si ccpvtz omega dependence}, we report the wall time for initialization and for per-iteration J-build with a (2,2,2) supercell of the Si crystal at $\Gamma$ point with varied $\omega$.
We also include time-to-solution, which consists of $8$ SCF iterations in total (obtained from the actual SCF runs in QCPBC), as a function of $\omega$.
The time-to-solution increases significantly when the workload between short-range GTO-DF and long-range/full-range PW-DF is not adequately balanced.
The optimal choice for $\omega$ depends on various factors, including the optimization level for real/reciprocal space kernels, lattice packing, and the diffuseness of the basis set.
As our short-range contributions require further optimization, we defer automatic determination of the optimal $\omega$ for later work and instead provide recommendations in the conclusion.
We anticipate that the optimal $\omega$ at the $\Gamma$ point would also be the optimal one for $\mathbf k$-point calculations because many time-consuming intermediate contraction steps are independent of $N_k$.

\renewcommand{\arraystretch}{1.3}
\begin{table*}[t] 
    \centering
    \begin{tabular}{ C{3cm} C{1cm} C{1cm} C{1cm} C{2cm} C{1.5cm} C{1.5cm} C{1.5cm} C{1.5cm} }
    \hline\hline
    System & $N_{\text{bsf}}$ & $N_{\text{aux}}$ & $N_{\text{elec}}$ & $N_k$ & QCPBC init & QCPBC per iter & QCPBC RI-J & PySCF RI-J \\ 
    \hline\hline
    \multirow{5}{*}{Benzene} & \multirow{5}{*}{456} & \multirow{5}{*}{2616} & \multirow{5}{*}{168} & \text{$1\times1\times1$} & 21.55 & 1.04 & 22.60 & 22.86   \\ 
                             &                      &  &  & \text{$2\times2\times2$} & 26.74 & 1.87 & 28.61 & 162.56  \\
                             &                      &  &  & \text{$3\times3\times3$} & 30.13 & 1.95 & 32.08 & 840.61 \\ 
                             &                      &  &  & \text{$4\times4\times4$} & 30.77 & 2.34 & 33.11 & 2153.08 \\ 
                             &                      &  &  & \text{$10\times10\times10$} & 32.21 & 9.97 & 42.18 & -       \\
    \hline\hline
    \multirow{5}{*}{Si (2,2,2)} & \multirow{5}{*}{544}  & \multirow{5}{*}{1936} & \multirow{5}{*}{224}  & \text{$1\times1\times1$} & 26.43 & 1.37 & 27.80 & 31.97    \\ 
                                &  &  &     & \text{$2\times2\times2$} & 32.49 & 2.11 & 34.60  & 176.17  \\
                                &  &  &     & \text{$3\times3\times3$} & 37.44 & 2.57 & 40.01 & 728.17  \\
                                &  &  &     & \text{$4\times4\times4$} & 38.33 & 3.01 & 41.34 & 1998.42 \\
                                &  &  &     & \text{$10\times10\times10$} & 40.16 & 11.75 & 51.91 & -      \\
    \hline\hline
    MgO(001) + CO  & 1052 & 3458 & 654 & \text{$1 \times 1 \times 1$} & 21.49 & 7.31 & 28.8 & 195.64 \\
    \hline\hline
    \end{tabular}
    \caption{Total wall time (seconds) to generate the Coulomb matrix of benzene crystal, Si (2,2,2), and MgO(001)+CO using \texttt{QCPBC} and \texttt{PySCF}. 
    We also report the timing breakdown of the \texttt{QCPBC} total time into initialization and per-iteration time.
    $N_\text{bsf}$, $N_\text{aux}$, and $N_\text{elec}$ are given for the unit cell.
    }
    \label{tab: timing against PySCF}
\end{table*}

\subsection{Timing benchmark against PySCF} \label{sec:timing against PySCF}

To assess the efficiency of our implementation, we provide some timing benchmarks against the all-electron calculation in \texttt{PySCF} \cite{sun2020recent} through RSDF. \cite{ye_fast_2021} 
The integral precision was set to $10^{-10}$ for \texttt{PySCF}.
In \texttt{QCPBC}, the timing was obtained using default screening settings specified at the beginning of \cref{sec:numerical section}.
The \texttt{PySCF} J-build time includes both the initialization time, which involves generating two-center and three-center integrals used in RSDF, and the actual J-build time that contracts the density matrix and integrals. \cite{ye_fast_2021}
Similarly to the \texttt{PySCF} RI-J time, the \texttt{QCPBC} RI-J time consists of initialization time to generate the necessary two-electron integrals (both CD/CC types) and one average time per iteration for calculating the Coulomb matrix.

For \texttt{QCPBC} J-build time, we use $\omega = 0.4$ for (2,2,2) Si supercell, which is optimal in ~\cref{fig:Si ccpvtz omega dependence}, with cc-pVTZ basis and MgO(001) + CO system with def2-SVP basis, and $\omega = 0.5$ for benzene crystal with cc-pVDZ basis.
Density classification (compact/diffuse) is turned on in all calculations. 
For \texttt{PySCF}, the default $\omega$, which varies depending on the system and $N_k$, was employed.
We observe that \texttt{PySCF}'s choice of range-separated parameter $\omega$ for systems presented here is typically less than $0.3$, which might come from balancing the cost of short-range integral generation and the expensive long-range AFT evaluation.
Our work here enables the fast on-the-fly evaluation of the long-range contribution, which makes the optimal $\omega$ larger and explains the choice of $\omega$ above.
Because the primitive GTOs in correlation-consistent basis sets can vary slightly across different sources, we ensured that we used identical cc-pVXZ (X = D, T) primary and auxiliary basis sets for all calculations to ensure a fair comparison.
For example, we use the \texttt{PySCF}'s default cc-pVTZ basis set for the Si crystal in the timing benchmark here. This basis set has fewer diffuse primitives for contracted S/P shells than the one from Ref.~\citenum{pritchard2019new} used for generating \cref{fig:Si ccpvtz omega dependence}.
Structural details about the benzene crystal and MgO(001) + CO are presented in \cref{sec:benzene crystal} and \cref{sec:MgO + CO}, respectively.

In ~\cref{tab: timing against PySCF}, we compare the total \texttt{QCPBC} time required for generating the Coulomb matrix against that of \texttt{PySCF}. 
We observe that the two programs achieve relatively similar performance for $\Gamma$-point calculations for both the benzene crystal and the (2,2,2) Si supercell.
For MgO(001) + CO, the \texttt{PySCF} RI-J time is significantly longer, primarily due to the evaluation of the AFT contribution.
Our fast per-iteration time is encouraging for our approach, as we perform the shell pair lattice summation on the fly for PW density-fitted parts, which one might suspect would be slower than approaches that store density-fitted integrals, such as RSDF.
Our efficient PW-DF introduces minimal computational overhead compared to reusing the pre-calculated density-fitted integral, as implemented in $\texttt{PySCF}$.

As we increase the number of $\mathbf k$-points, our implementation outperforms RSDF by orders of magnitude. 
Even up to $10\times10\times10$ $\mathbf k$-mesh, our Coulomb construction time only increases by a small factor compared to the $\Gamma$-point calculation, whereas $\texttt{PySCF}$ shows a relatively steep cost increase with $N_k$ due to the $\mathcal O(N_k^{1-2})$ scaling steps associated with \cref{eq:naive ft}.
We attribute the much softer $N_k$-dependence of our method to our real-space approach to the short-range integrals and the efficient evaluation of the PW-DF contributions. 
Furthermore, we anticipate that the memory requirement for $\mathbf k$-space integrals in RSDF will be significantly larger than that of our BvK format integral, which will add additional cost if I/O is required.

\subsection{Application to benzene crystal} \label{sec:benzene crystal}

The unit cell of the benzene crystal contains four benzene molecules, each comprising 42 electrons.
If the BvK representation were not used, calculations with atom-centered basis sets could easily be constrained to a small $\mathbf k$-mesh due to the steep computational cost associated with $N_k$.
Here, we obtain the cohesive energy of the molecular benzene crystal at the thermodynamic limit using the 138 K lattice geometry. \cite{bacon_crystallographic_1997}
The largest calculation here employs the cc-pVQZ basis set with $4\times4\times4$ $\mathbf k$-point mesh, which represents 2,040 basis functions per $\mathbf k$-point. 
The cohesive energy for molecular crystals is given by
\begin{equation} \label{eq:benzene cohesive energy}
    E_{\text{coh}} = \frac{E_\text{crystal} - N_{\text{mol}} \cdot E_\text{mol}}{N_{\text{mol}}},
\end{equation}
where $E_\text{crystal}$ is the energy of the benzene crystal converged to the thermodynamic limit,
$E_\text{mol}$ is the energy of a molecular fragment in the unit cell, and $N_{\text{mol}}$ is the number of molecules in the unit cell (in this case, 4).
The counterpoise correction was carried out by including ghost atomic shells in a (2,2,2) supercell. 

In \cref{tab: PBE-D3 cohesive energy for benzene}, we observe that the cohesive energy has converged to the thermodynamic limit with $3\times3\times3$ $\mathbf k$-mesh, consistent with a previous study. \cite{rettig2023even}
The DFT-D3 dispersion treatment with Becke-Johnson (BJ)\cite{grimme2011effect} damping is utilized in the calculation.
The cohesive energy obtained here aligns qualitatively with a previous study on using dispersion corrections for benzene and the X23 dataset using PAW potentials. \cite{moellmann2014dft}
In their study, the cohesive energy of benzene, which includes pairwise DFT-D3 dispersion energy and the Becke–Johnson (BJ) damping, is reported to be $-54.51$ kJ/mol.
Interestingly, this DFT result is in qualitative agreement with the theoretical best estimate (TBE) of $-54.58$ kJ/mol. \cite{yang2014ab}

\begin{table} [H]
\captionsetup{type=table}
\captionof{table}{Cohesive energy (kJ/mol) for benzene crystal with PBE-D3(BJ) functional.}
\begin{tabular}{ C{2cm} C{2cm} C{2cm} C{2cm} }
\hline
\hline
$N_k$ & DZ & TZ & QZ \\ 
\hline
\hline
\text{$1\times1\times1$} & -54.86 & -55.14 & -55.35 \\ 
\text{$2\times2\times2$} & -53.76 & -53.84 & -54.03 \\
\text{$3\times3\times3$} & -53.76 & -53.85 & -54.04 \\ 
\text{$4\times4\times4$} & -53.76 & -53.85 & -54.04 \\ 
\hline
\hline
\label{tab: PBE-D3 cohesive energy for benzene}
\end{tabular} 
\end{table}

\subsection{Application to CO on MgO(001)}  \label{sec:MgO + CO}
The adsorption of CO on the MgO (001) surface problem has been examined extensively with both DFT \cite{valero2008good} and wavefunction methods. \cite{shi2023many, alessio2018chemically, mitra2022periodic, staemmler2011method}
Obtaining a converged interaction energy at low CO coverage requires systematic studies with respect to slab sizes and thickness.
This represents a test for a relatively large $\Gamma$-point calculation.

Here, we employ a two-layer 4$\times$4 supercell MgO slab, which contains a total of 64 atoms; this setup has been demonstrated to be suitable for calculating adsorption energies. \cite{ye2024adsorption, mitra2022periodic}
The equilibrium geometry for the CO-MgO composite system, which was optimized using the PBE-D3 functional, was taken from Ref.~\citenum{ye2024adsorption}.
The CO bond length is 1.140 \AA, and the distance from the CO to the MgO surface is 2.377 \AA.
All relative energies were counterpoise corrected without relaxing the geometry of each subsystem.
Therefore, the interaction energy reported here does not account for the geometry distortion, which is around $1.0$ kJ/mol.~\cite{boese2013accurate, alessio2018chemically}

In \cref{tab: PBE for CO on MgO}, we reported the adsorption energy with various def2 basis sets \cite{weigend2005balanced} at the $\Gamma$-point.
The universal auxiliary Coulomb fitting basis set is used throughout. \cite{weigend2006accurate} 
We observed that RI-J energies are sometimes numerically unstable for larger (TZ/QZ) basis sets with shell pair classification. 
We attribute this to the DF error, as employing a larger DF basis set (e.g., cc-pVQZ-jkfit) eliminates the instability.
We also found that the instability is no longer present when shell-pair classification is omitted (i.e., classifying every shell pair as a compact one).
The def2-universal Coulomb fitting basis set was still employed, but the shell pair classification step was disabled for this example.

Compared to the experimental value of $-20.0$ kJ/mol, the dispersion-corrected PBE functional significantly overestimates the binding energy by around 10 kJ/mol.
This finding is in agreement with the overbinding observation with PBE-D3(0) functional in Ref. \citenum{ye2024adsorption}, which reported the adsorption energy of $-29.5$ kJ/mol using the projector augmented wave method in Quantum Espresso.~\cite{carnimeo2023quantum}

\begin{table}[H]
\captionsetup{type=table}
\captionof{table}{Binding energy (kJ/mol) for adsorption of CO on MgO (001) surface.}
\begin{tabular}{C{2.0cm} C{2.0cm} C{2.0cm} C{2.0cm} }
\hline
\hline
Functional & def2-SVP & def2-TZVP & def2-QZVP \\ 
\hline
\hline
PBE-D3(BJ) &-31.12  & -32.17 & -32.08 \\ 
PBE-D3(0) & -29.29  & -30.34 & -30.25 \\ 
\hline
\hline
\label{tab: PBE for CO on MgO}
\end{tabular}
\end{table}

\section{Conclusions} \label{sec:conclusion}

In summary, we have presented an efficient scheme for evaluating the periodic all-electron Coulomb matrix.
Our short-range GTO-DF is implemented in real space within the Born–von Kármán supercell formalism, exploiting the shell-pair sparsity. 
We also introduce an efficient algorithm to accelerate the PW-DF used in the Ewald summation, allowing for the fast on-the-fly evaluation of long-range and diffuse density contributions to the Coulomb matrix.
Our algorithm controls the work balance between GTO-DF and PW-DF by a single range-separation parameter, $\omega$.
Additionally, compact/diffuse classification can also be employed to further balance the computational cost.
While a broader range of tests is needed, we recommend $\omega$ = 0.4 -- 0.5 for typical solids.
If a large vacuum is present, such as the one in slab calculations, a smaller $\omega$ = 0.3 -- 0.4 is recommended to limit the number of PWs.
Overall, our work demonstrates that semilocal density functional calculations can be carried out computationally efficiently using the traditional molecular Gaussian basis sets.

Several directions exist for improving upon the algorithm presented here. 
The short-range component can be evaluated on-the-fly using the J-engine technique, \cite{white1996aj,wang2024fast} thereby eliminating the memory bottleneck associated with storing the short-range integrals. 
This approach also eliminates the cubic-scaling bottleneck in RI-J, resulting in an overall quadratic-scaling algorithm.
In addition to algorithmic enhancements, incorporating space-group symmetry \cite{cao2025applying} into the present real-space formalism could reduce the number of significant integrals. 
Furthermore, building upon our group's recent works, we will extend our algorithms to enable efficient all-electron hybrid DFT,~\cite{rettig2023even} correlated wavefunction,~\cite{chen2025regularized} and constant-potential DFT calculations.~\cite{ni2025gaussian}

\section*{Data Availability Statement}
The data that support the findings of this study can be found at Ref.~\citenum{mydataset}.

\section*{Code availability}
Plotting codes used in this study are available at Ref.~\citenum{mydataset}.

\begin{acknowledgments}
This work was supported by the startup fund provided by Harvard University.
A.R. was partly supported by the Harvard Quantum Initiative prize postdoctoral fellowship.
H.Q.D. thanks all members of the Lee group for their support. 
We thank Fionn Malone, Hong-Zhou Ye, and Tim Berkelbach for valuable discussions.
The simulation was performed on the FASRC cluster, supported by the FAS Division of Science Research Computing Group at Harvard University.
This work also used the Delta system at the National Center for Supercomputing Applications through allocation CHE250005 from the Advanced Cyber infrastructure Coordination Ecosystem: Services \& Support (ACCESS) program, which is supported by National Science Foundation grants \#2138259, \#2138286, \#2138307, \#2137603, and \#2138296.
\end{acknowledgments}

\appendix

\section{Hermite Polynomials} \label{sec:hermite polynomials}

As shown in the main text, the angular component of the Fourier Transform for a Cartesian Gaussian function with angular momentum $l$ is the physicist Hermite polynomial $H_l(G)$. These polynomials satisfy the following recurrence relation
\begin{equation}
    H_n(G) = 2G \cdot H_{n-1}(G) - 2(n-1) \cdot H_{n-2}(G),
\end{equation}
with base cases given by
\begin{align*}
    H_0(G) &= 1, \\
    H_1(G) &= 2G.
\end{align*}

\section{Cartesian coordinates and relative coordinates} \label{sec:cart coord vs rel coord}

The integration grid for the Coulomb matrix in this work is sampled uniformly along either the direct lattice vector or the reciprocal lattice vector.
For general unit cell geometry, one can transform the basis between relative coordinates and Cartesian coordinates as follows
\begin{equation} \label{eq:transform rcart to rrel}
    [x,y,z]^{T} = \mathbf{A} \cdot [i,j,k]^{T} 
\end{equation}
\begin{equation} \label{eq:transform rrel to rcart}
    [i,j,k]^{T} = \frac{1}{2\pi} \mathbf{B}^{T} \cdot [x,y,z]^{T}
\end{equation}
\begin{equation} \label{eq:transform Gcart to Grel}
    \left[G_x, G_y, G_z\right]^T=\mathbf{B} \cdot\left[G_i, G_j, G_k\right]^T
\end{equation}
\begin{equation} \label{eq:transform from Grel to Gcart}
    \left[G_i, G_j, G_k\right]^T= \frac{1}{2\pi} \mathbf{A}^{T} \cdot\left[G_x, G_y, G_z\right]^T
\end{equation}
where $\mathbf{A} = [\mathbf{a}_i \; \mathbf{a}_j \; \mathbf{a}_k]$ and $\mathbf{B} = [\mathbf{b}_i \; \mathbf{b}_j \; \mathbf{b}_k]$ have columns corresponding to the primitive direct lattice vector and primitive reciprocal lattice vector, respectively. These transformation matrices satisfy 
\begin{equation}
    \mathbf{A} \cdot \mathbf{B}^{\mathrm{T}}=\mathbf{B}^{\mathrm{T}} \cdot \mathbf{A}=2 \pi \cdot \mathbf{I}.
\end{equation}
For an orthorhombic unit cell, all the direct (reciprocal) lattice vectors are mutually orthogonal. We have
\begin{equation} \label{eq:ortho condition}
    \mathbf{A}_{\text{ortho}} \cdot \mathbf{A}^{T}_{\text{ortho}} = \Lambda,
\end{equation}
where $\Lambda$ is the diagonal matrix with entries corresponding to the square norm of primitive lattice vectors.

\section{Expansion of angular prefactor in relative coordinate}
\label{sec:appendix_prefactor}
Given the basis transformation ~\cref{eq:transform Gcart to Grel}, we can expand the angular prefactor of the Fourier Transformed Gaussian functions with the polynomial triplet $(l_x, l_y, l_z)$ to the relative coordinate as follows

\begin{equation} \label{eq:transform cart angular factor to relative angular factor}
\begin{aligned}
G_x^{l_x} G_y^{l_y} G_z^{l_z} = \sum^{\substack{
l_{x i} + l_{x j} + l_{x k} = l_x \\
l_{y i} + l_{y j} + l_{y k} = l_y \\
l_{z i} + l_{z j} + l_{z k} = l_z
}}_{\substack{
l_{x i}, l_{x j}, l_{x k} \\
l_{y i}, l_{y j}, l_{y k} \\
l_{z i}, l_{z j}, l_{z k}
}}
&\binom{l_{x}}{l_{x i}\,l_{x j}\,l_{x k}} \binom{l_{y}}{l_{y i}\, l_{y j}\,l_{y k}} \binom{l_{z}}{l_{z i}\,l_{z j}\,l_{z k}}   \\
& B_{11}^{l_{x i}} B_{12}^{l_{x j}} B_{13}^{l_{x k}}
B_{21}^{l_{y i}} B_{22}^{l_{y j}} B_{23}^{l_{y k}} B_{31}^{l_{z i}} B_{32}^{l_{z j}} B_{33}^{l_{z k}} \\
&  G_i^{l_{xi} + l_{yi} + l_{zi}} G_j^{l_{xj} + l_{yj} + l_{zj}} G_k^{l_{xk} + l_{yk} + l_{zk}},
\end{aligned}
\end{equation}
where the multinomial coefficients are 
\begin{equation} \label{eq:multinomial equation}
    \binom{l_{x}}{l_{x i}\, l_{x j} \,l_{x k}} = \frac{l_x!}{l_{xi}! l_{xj}! l_{xk}!}.
\end{equation}

\section{Evaluation of planewave fitted density } \label{sec:eval of density on reciprocal space}

In this section, we focus on the relevant transformation of a given primitive shell pair (ab) within the list of contracted shell pairs $|\mu \mathbf{0} \nu \mathbf{n})$. We drop the lattice index $\mathbf{n}$ for clarity. 
The contribution of a primitive pair to the PW fitted density is given by
\begin{equation}
\begin{aligned}
    \rho(\mathbf{G}) \mathrel{+}= &\sum_{\substack{ l_{x_a}, l_{y_a}, l_{z_a} \\ 
                                    l_{x_b}, l_{y_b}, l_{z_b} } }
                        D_{l_{x_a}, l_{y_a}, l_{z_a}, l_{x_b}, l_{y_b}, l_{z_b}} \\
                        &\int (x - X_{a})^{l_{x_a}} (x - X_{b})^{l_{x_b}} e^{-\alpha_{a} |x - X_{a}|^2 - \alpha_{b} |x - X_{b}|^2 -i G_x x} dx \\
                        &\int (y - Y_{a})^{l_{y_a}} (y - Y_{b})^{l_{y_b}} e^{-\alpha_{a} |y - Y_{a}|^2 - \alpha_{b} |y - Y_{b}|^2 - i G_y y} dy \\
                        &\int (z - Z_{a})^{l_{z_a}} (z - Z_{b})^{l_{y_b}} e^{-\alpha_{a} |z - Z_{a}|^2 - \alpha_{b} |z - Z_{b}|^2 -i G_z z} dz,
\end{aligned}
\end{equation}
where $l_{x_a}, l_{y_a}, l_{z_a} \in \{ 0,..., l_{a}^{\max} \}$, and $l_{x_b}, l_{y_b}, l_{z_b} \in \{  0,..., l_{b}^{\max} \}$.
Using the Gaussian product theorem on each Cartesian component, we can convert the product of polynomials into a new series of polynomials with the highest order of $l_p^{\max} = l_{a}^{\max} + l_{b}^{\max}$. 
The unfolded density matrix is converted with the corresponding prefactor to match this new polynomial series. We have
\begin{equation}
\begin{aligned}
    \rho(\mathbf{G}) \mathrel{+}= &\sum_{ l_{x_p}, l_{y_p}, l_{z_p} } D_{l_{x_p}, l_{y_p}, l_{z_p}} \\
                    &\int (x - X_p)^{l_{x_p}} e^{-\alpha_{p} |x - X_{p}|^2 -i G_x x} dx \\
                    &\int (y - Y_p)^{l_{y_p}} e^{-\alpha_{p} |y - Y_{p}|^2 - i G_y y} dy \\
                    &\int (z - Z_p)^{l_{z_p}} e^{-\alpha_{p} |z - Z_{p}|^2 -i G_z z} dz,
\end{aligned}
\end{equation}
where $l_{x_p}, l_{y_p}, l_{z_p} \in \{ 0,..., l_{a}^{\max} + l_{b}^{\max} \}$.
We now apply the Analytical Fourier Transform (\cref{eq:FT of cart Gaussians}) to each polynomial order and then expand them as polynomials of $G_x, G_y$, and $G_z$; again, all the prefactors are used to transform the density matrix,
\begin{equation}
\begin{aligned}
    \rho(\mathbf{G}) \mathrel{+}= \sum_{l_{G_x}, l_{G_y}, l_{G_z}} & D_{l_{G_x}, l_{G_y}, l_{G_z}} G_x^{l_{G_x}} G_y^{l_{G_y}} G_z^{l_{G_z}} \\
    &e^{-|\mathbf{G}|^2/4{\alpha_p}} e^{-i (G_x X_p + G_y Y_p + G_z Z_p)}
\end{aligned}
\end{equation}
where $l_{G_{x}}, l_{G_{y}}, l_{G_{z}} \in \{ 0,..., l_{a}^{\max} + l_{b}^{\max} \}$. \\

If the unit cell is not orthorhombic, we use \cref{eq:transform cart angular factor to relative angular factor} to expand the angular factor in relative coordinates. With the prefactor absorbed into the density matrix, we obtain
\begin{equation}
\begin{aligned}
    \rho(\mathbf{G}) \mathrel{+}= &\sum_{l_{G_i}, l_{G_j}, l_{G_k}} D_{l_{G_i}, l_{G_j}, l_{G_k}} G_i^{l_{G_i}} G_j^{l_{G_j}} G_k^{l_{G_k}} \\
    & E(G_i, G_j) E(G_j, G_k) E(G_k, G_i) e^{-i 2\pi \cdot \tilde{\mathbf{G}} \tilde{\mathbf{R}}_p},
\end{aligned}
\end{equation}
where $\tilde{\mathbf{R}}_p = [I_p, J_p, K_p]$ is the relative coordinate of the shell pair center, and the non-separable exponential term $E(G_i, G_j)$ is defined as in \cref{eq:non-separable exponential term}.

\section{Evaluation of density on real space grid (GPW)} \label{sec:accelerate rho build GPW}
Evaluation of the electron density, $\rho(\mathbf{r})$, can also be done on the fly by looping over pairs of primitive Gaussian functions and separating the evaluation into each grid dimension. \texttt{CP2K} has employed this strategy to speed up calculations \cite{vandevondele2005quickstep} for many years, with great effect, by designing basis sets that use only a handful of unique primitive Gaussians. \\

The density is constructed within the BvK supercell formalism as follows
\begin{equation}
\rho(\mathbf{r}) = \sum_{\mathbf{n},\mathbf{m}} \sum_{\mu \nu} \phi_\mu^\mathbf{m}(\mathbf{r})\phi_\nu^\mathbf{n}(\mathbf{r}) D_{\mu\nu}^{\mathbf{m} \mathbf{n} },
\end{equation}
where $\mathbf{r}$ is sampled uniformly inside the central unit cell.

We fix the contribution from a pair of shells $\mu \mathbf{m}$ and $\nu \mathbf{n}$ for easier manipulation.
The basis functions $\phi_\mu(\mathbf{r})$  are then further broken down into the set of primitive Gaussian-type orbitals $\chi_{a}(\mathbf{r})$ within the contracted shell pair, similar to PW-fitted density,
\begin{equation}
\rho(\mathbf{r}) \mathrel{+} = 
D^{\mathbf{m}\mathbf{n}}_{\mu\nu} \sum_{ab}C_{\mu a} C_{\nu b} \cdot \chi_{a}(\mathbf{r} - \mathbf{R}_{\mathbf{m}}) \chi_{b}(\mathbf{r} - \mathbf{R}_{\mathbf{n}}).
\end{equation}

The subblock of the density matrix is multiplied by the contraction coefficients $(C_{\mu a}, C_{\nu b})$ before it is unfolded. The contribution of the primitive pair to the density follows
\begin{equation}
\begin{aligned}
\label{eq:rho_shellpair}
\rho(\mathbf{r})  \mathrel{+} = &\sum_{\substack{ l_{x_a}, l_{y_a}, l_{z_a}  \\l_{x_b}, l_{y_b}, l_{z_b}}} D_{\substack{ l_{x_a}, l_{y_a}, l_{z_a}, l_{x_b}, l_{y_b}, l_{z_b}}} \\
&(x-X_a)^{l_{x_a}} (x-X_b)^{l_{x_b}} e^{-\alpha_{a}|x-X_a|^2} e^{-\alpha_{b}|x-X_b|^2}  \\
&(y-Y_a)^{l_{y_a}} (y-Y_b)^{l_{y_b}} e^{-\alpha_{a}|y-Y_a|^2} e^{-\alpha_{b}|y-Y_b|^2}  \\
&(z-Z_a)^{l_{z_a}} (z-Z_b)^{l_{z_b}} e^{-\alpha_{a}|z-Z_a|^2} e^{-\alpha_{b}|z-Z_b|^2},  \\
\end{aligned}
\end{equation}
where $l_{x_a}, l_{y_a}, l_{z_a} \in \{ 0,..., l_{a}^{\max} \}$, and $l_{x_a}, l_{y_a}, l_{z_a} \in \{  0,..., l_{b}^{\max} \}$. This allows for effective screening - very compact primitive Gaussian functions will only be evaluated on a small set of grid points and the lattice sum $\mathbf{n}, \mathbf{m}$ will be quickly truncated.
We note that in our implementation, instead of shifting both shells $\mu$ and $\nu$, we perform the lattice sum over shell $\nu$ and the grid points in the central unit cell.

\subsection{Orthorhombic unit cells}
In the case of orthorhombic unit cells, the contraction in ~\cref{eq:rho_shellpair} may be evaluated directly as the $x,y,z$ coordinates are themselves the coordinates of the lattice grid. 
Two speedups arise from the separability of the Gaussian functions between the $x$, $y$, and $z$ dimensions. First, each one-dimensional contribution can be evaluated separately in the whole space, then wrapped back to the central unit cell before being contracted with the other dimensions. Second, the density matrix $D$ may be separately contracted with each dimension of the whole shell pair: the $D_{l_{x_a}, l_{y_a}, l_{z_a}, l_{x_b}, l_{y_b}, l_{z_b}}$ submatrix may be first contracted over $l_{x_a}$ and $l_{x_b}$ - which scales only with the number of grid points in the $x$ dimension, $N_{G_x}$. 
Then, the resulting tensor may be contracted over $l_{y_a}$ and $l_{y_b}$ which scales with $N_{G_x}N_{G_y}$, and finally over $l_{z_a}$ and $l_{z_b}$ which will be the only step that scales with $N_G$.

\subsection{Non-orthorhombic unit cells}
In the case of non-orthorhombic unit cells, the basis functions do not separate along the grid dimensions, requiring an additional transformation. The angular components of the basis functions may be transformed to a commensurate form separated in the grid dimensions $\{i,j,k\}$, using the strategy outlined in~\cref{sec:appendix_prefactor}
\begin{align}
(x&-X_a)^{l_{x_a}}(y-Y_a)^{l_{y_a}}(z-Z_a)^{l_{z_a}} = \nonumber \\
& \sum^{\substack{
l_{x i_a} + l_{x j_a} + l_{x k_a} = l_{x_a} \\
l_{y i_a} + l_{y j_a} + l_{y k_a} = l_{y_a} \\
l_{z i_a} + l_{z j_a} + l_{z k_a} = l_{z_a}
}}_{\substack{
l_{x i_a}, l_{x j_a}, l_{x k_a} \\
l_{y i_a}, l_{y j_a}, l_{y k_a} \\
l_{z i_a}, l_{z j_a}, l_{z k_a}
}} d_{\substack{
l_{x i_a}, l_{x j_a}, l_{x k_a} \\
l_{y i_a}, l_{y j_a}, l_{y k_a} \\
l_{z i_a}, l_{z j_a}, l_{z k_a}
}} (i-I_a)^{l_{i_a}}  (j-J_a)^{l_{j_a} }  ( k-K_a)^{l_{k_a}},
\end{align}
where the coefficients $d$, are given by
\begin{align}
d_{\substack{
l_{x i_a}, l_{x j_a}, l_{x k_a} \\
l_{y i_a}, l_{y j_a}, l_{y k_a} \\
l_{z i_a}, l_{z j_a}, l_{z k_a}
}} = &\binom{l_{x_a}}{l_{x i_a}\,l_{x j_a}\,l_{x k_a}}  \binom{l_{y_a}}{l_{y i_a}\,l_{y j_a}\,l_{y k_a}} \binom{l_{z_a}}{l_{z i_a}\,l_{z j_a}\,l_{z k_a}} \nonumber \\
\times&  A_{11}^{l_{xi_a}} A_{12}^{l_{xj_a}} A_{13}^{l_{xk_a}} 
         A_{21}^{l_{yi_a}} A_{22}^{l_{yj_a}} A_{23}^{l_{yk_a}}
         A_{31}^{l_{zi_a}} A_{32}^{l_{zj_a}} A_{33}^{k_{zk_a}}.
\end{align}

The $d$ coefficients can be contracted with the density matrix in Cartesian coordinates to yield a density matrix resolved in angular momentum components $(l_{i_a}, l_{j_a}, l_{k_a})$
\begin{equation}
\begin{aligned}
D_{\substack{ l_{i_a} l_{j_a} l_{k_a}  \\l_{x_b} l_{y_b} l_{z_b}}} = \sum^{\substack{
l_{x i_a} + l_{y i_a} + l_{z i_a} = l_{i_a} \\
l_{x j_a} + l_{y j_a} + l_{z j_a} = l_{j_a} \\
l_{x k_a} + l_{y j_a} + l_{z k_a} = l_{k_a}
}}_{\substack{
l_{x i_a}, l_{x j_a}, l_{x k_a} \\
l_{y i_a}, l_{y j_a}, l_{y k_a} \\
l_{z i_a}, l_{z j_a}, l_{z k_a}
}}  D_{\substack{ l_{x_a} l_{y_a} l_{z_a}  \\l_{x_b} l_{y_b} l_{z_b}}} \cdot d_{\substack{
l_{x i_a}, l_{x j_a}, l_{x k_a} \\
l_{y i_a}, l_{y j_a}, l_{y k_a} \\
l_{z i_a}, l_{z j_a}, l_{z k_a}
}}  \\
\end{aligned}
\end{equation}
Note that the above $D$ matrix is only half-transformed. An identical transformation must also be performed on the ket, $b$, dimensions.

This density matrix may now be used in the evaluation of $\rho(\mathbf{r})$, giving an expression similar to the orthorhombic case. However, there is the added complication that the Gaussian functions do not entirely separate and must be written as the product of two-dimensional functionals,
\begin{equation}
\begin{aligned}
\label{eq:rho_shellpair_nonortho}
\rho(\mathbf{r}) \mathrel{+}= \sum_{\substack{ l_{i_a} l_{j_a} l_{k_a}  \\l_{i_b} l_{j_b} l_{k_b}}}& D_{l_{i_a}, l_{j_a}, l_{k_a}, l_{i_b}, l_{j_b}, l_{k_b}} \\
& (i-I_a)^{l_{i_a}} (i-I_b)^{l_{i_b}}   \\
& (j-J_a)^{l_{j_a}} (j-J_b)^{l_{j_b}}  \\
& (k-K_a)^{l_{k_a}} (k-K_b)^{l_{k_b}}  \\
& h_{a}(i,j) h_{a}(j,k) h_{a}(k,i)    \\ 
& h_{b}(i,j) h_{b}(j,k) h_{b}(k,i),    \\ 
\end{aligned}
\end{equation}
where
\begin{align}
h_{a}(i,j) = \text{exp}\bigg[-\alpha_{a}\big(\mathbf{a}_{i} \cdot \mathbf{a}_{i} (i-I_a)^2 + 2 \mathbf{a}_i \cdot \mathbf{a}_j (i-I_a)(j-J_a)\big) \bigg].
\end{align}

In our implementation, the product of $a$ and $b$ Gaussian functions in the shell pair is also rewritten as a single $p$ Gaussian function as in ~\cref{eq:Gaussian product of the radial decay}. This Gaussian is then transformed in a similar fashion, where all prefactors are used to transform the density matrix. The contribution to the density is given by
\begin{equation}
\begin{aligned}
\rho(\mathbf{r}) \mathrel{+}= \sum_{l_{i_p} l_{j_p} l_{k_p}}& D_{l_{i_p}, l_{j_p}, l_{k_p}}
(i-I_p)^{l_{i_p}} (j-J_p)^{l_{j_p}} (k-K_p)^{l_{k_p}}   \\
& h_{p}(i,j) h_{p}(j,k) h_{p}(k,i),    \\ 
\end{aligned}
\end{equation}
where $l_{i_p}, l_{j_p}, l_{k_p} \in \{0, \ldots, l_a^{\max} + l_b^{\max} \}$.

Non-orthorhombic unit cells are therefore slower to evaluate than orthorhombic unit cells due to the need for an extra transformation as well as the non-separability of the Gaussian functions.

\nocite{*}
\bibliography{main}
\clearpage
\onecolumngrid

\renewcommand\thesection{S\arabic{section}}
\renewcommand\thesubsection{\thesection.\arabic{subsection}}
\setcounter{section}{0}

\renewcommand{\theequation}{S\arabic{equation}}
\setcounter{equation}{0}
\renewcommand{\thefigure}{S\arabic{figure}}
\setcounter{figure}{0}
\renewcommand{\thetable}{S\arabic{table}}
\setcounter{table}{0}
\renewcommand{\thealgorithm}{S\arabic{algorithm}}
\setcounter{algorithm}{0}

\makeatletter

\renewcommand{\section}{\@ifstar\rtx@sectionstar\rtx@Supp@section}
\newcommand{\rtx@Supp@section}[1]{
  \refstepcounter{section}
  \par\vspace{2ex}
  {\normalfont\large\bfseries Supplementary Information \thesection.\ #1\par}
  \vspace{1ex}
  \addcontentsline{toc}{section}{Supplementary Information \thesection.\ #1}
}

\providecommand{\theHsection}{S\arabic{section}}
\providecommand{\theHsubsection}{S\arabic{section}.\arabic{subsection}}
\makeatother

\providecommand{\crefname}[3]{}
\providecommand{\Crefname}[3]{}
\crefname{section}{Supplementary Information}{Supplementary Information}
\Crefname{section}{Supplementary Information}{Supplementary Information}

\begin{center}
\textbf{\Large Supplementary Information: \\[0.5em] 
Efficient All-Electron Periodic Fourier-Transformed Coulomb Method}

\vspace{2em}

Hieu Q. Dinh\textsuperscript{1,†}, Adam Rettig\textsuperscript{1,†}, Xintian Feng\textsuperscript{2}, and Joonho Lee\textsuperscript{1,*} \\[1em]

\textsuperscript{1}Department of Chemistry and Chemical Biology, Harvard University, Cambridge, MA, USA \\
\textsuperscript{2}Q-Chem, Inc., 6601 Owens Drive, Suite 240, Pleasanton, California 94588, USA \\[1em]

\textsuperscript{†}These authors contributed equally to this work. \\
\textsuperscript{*}Email: joonholee@g.harvard.edu
\end{center}

\section{Numerical results and timing for LiF crystal} \label{sec:LiF SCF energies}

For LiF crystal, we employ def2 basis set series \cite{weigend2005balanced} with the universal auxiliary Coulomb fitting basis set. \cite{weigend2006accurate} 
This basis set series is more compact and computationally cheaper than the correlation-consistent one for DFT calculations.
The metric matrix $(P \tilde{\mathbf{0}}|Q \tilde{\mathbf{0}})_{\omega}$ is extremely ill-conditioned due to the Li diffuse auxiliary shells. 
To ensure numerical stability when solving linear equations, we remove two S shells with exponents of $0.0549139$ and $0.1054201$ from the Li auxiliary basis.

\begin{figure}[H]
    \centering
    \includegraphics[width=0.4\textwidth]{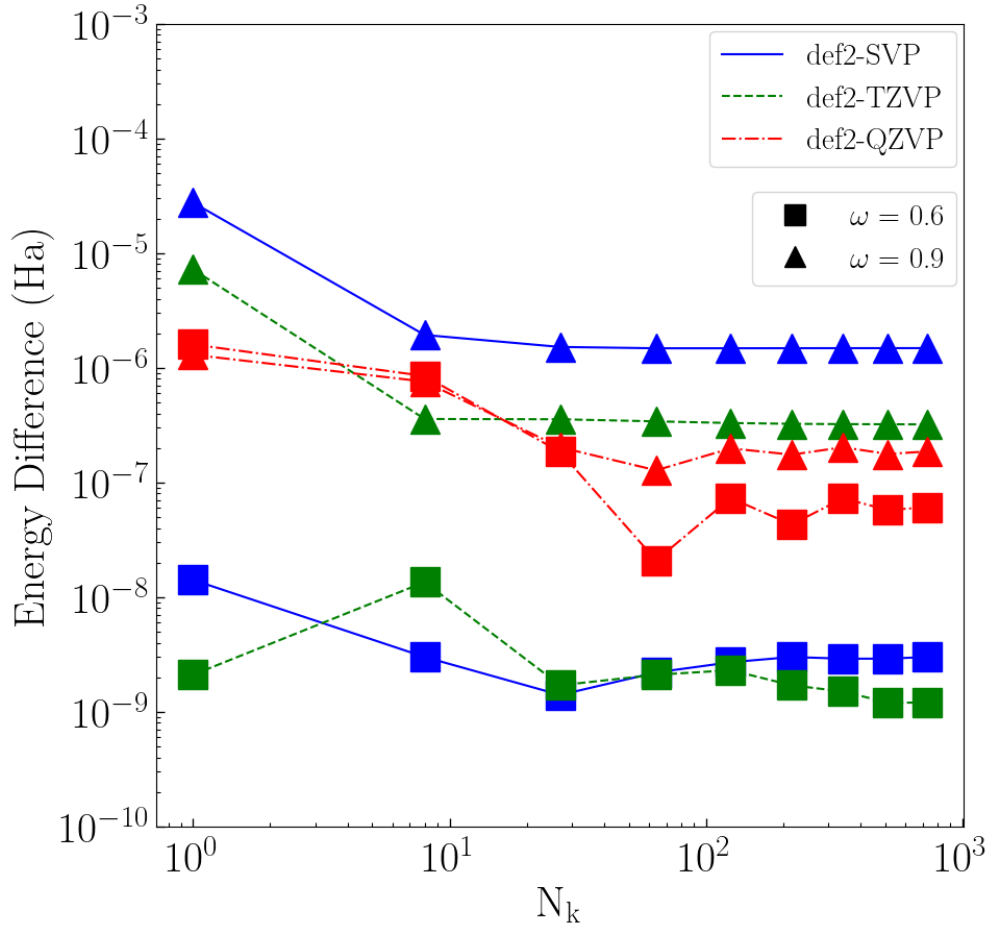}
    \caption{Difference in the converged SCF energies for the LiF crystal with def2-SVP, def2-TZVP, and def2-QZVP basis sets at various range-separated parameter values. The energy difference is reported as the absolute difference to the SCF energies obtained with $\omega = 0.3$. The number of sampled k-mesh is from $1\times1\times1$ to $9\times9\times9$.}
    \label{fig:lif energy (cd vs no cd)}
\end{figure}

In~\cref{fig:lif energy (cd vs no cd)}, we directly compare the converged energy at various $\omega$ values for LiF calculations against the one at $\omega =0.3$. 
We also observe that the change in energy is within the typical density fitting error.

\begin{figure}[H]
    \centering
    \includegraphics[width=0.6\linewidth]{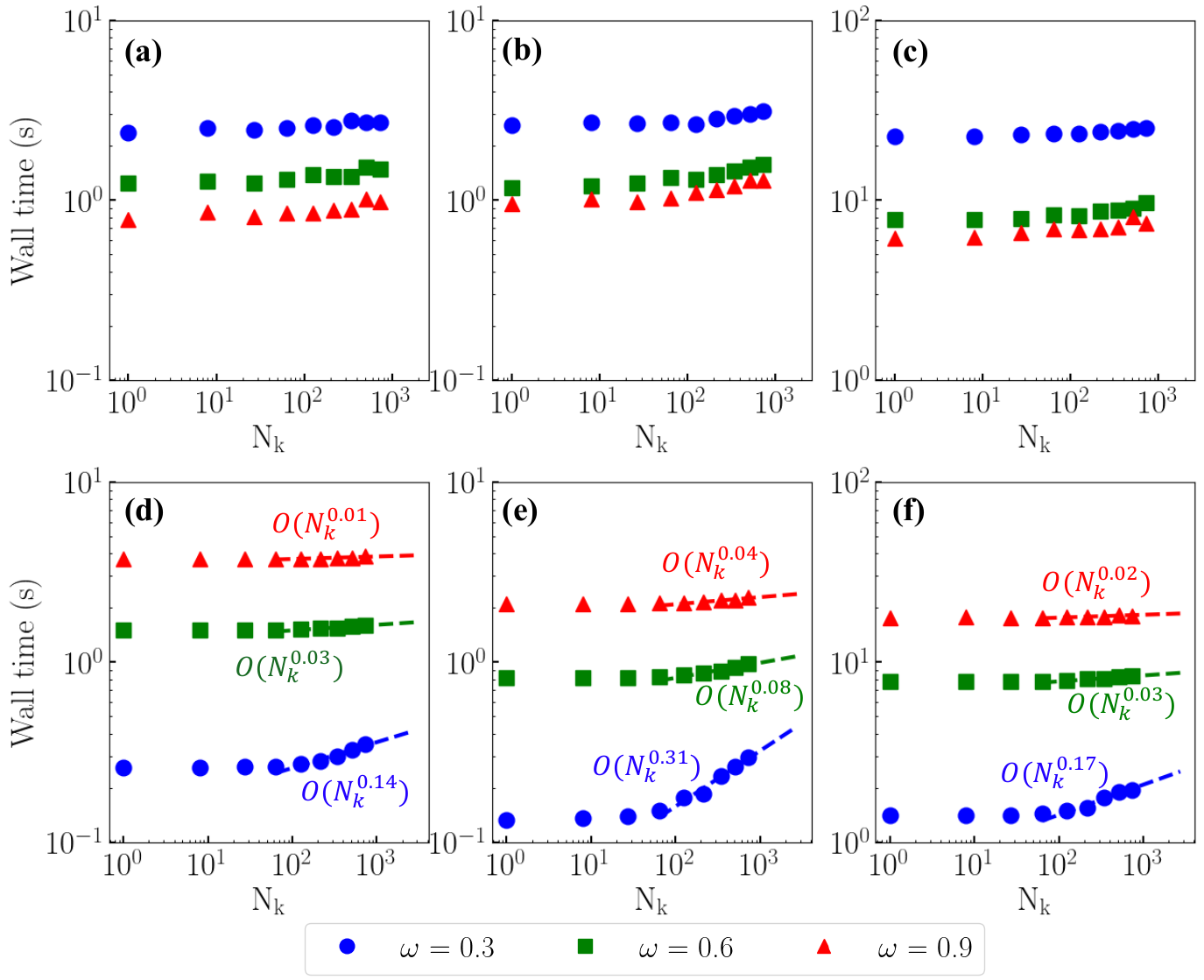}
    \caption{Initialization time at various range-separated parameter values $\omega$ for LiF crystal with \textbf{(a)} def2-SVP, \textbf{(b)} def2-TZVP, and \textbf{(c)} def2-QZVP basis set. 
    The initialization time consists of calculating the three-center integral and the two-center integral.
    Time per iteration for various range-separated parameter values is also included for the LiF crystal with \textbf{(d)} def2-SVP, \textbf{(e)} def2-TZVP, and \textbf{(f)} def2-QZVP basis sets.
    The dependence of the time per iteration on $N_k$ is fitted with the five largest $\mathbf k$-point meshes.
     The number of sampled $\mathbf k$-mesh is from $1\times1\times1$ to $9\times9\times9$.}
    \label{fig:lif crystal run time}
\end{figure}

In ~\cref{fig:lif crystal run time}, we observe that both initialization time and time per iteration show very weak dependence on $N_k$ even up to $9^3$ $\mathbf k$-point mesh.
The appearance of very diffuse primitive GTOs leads to many more significant shell pairs $(\mu \mathbf{0} \nu \mathbf{n})$, making the $N_k$-dependence for time per iteration much less than that of the Si crystal.
These observations are based on our analysis in the main text.

\newpage
\section{Extra data for Si crystal} \label{sec:extra data for Si}

\begin{figure}[H]
    \centering
    \includegraphics[width=0.5\linewidth]{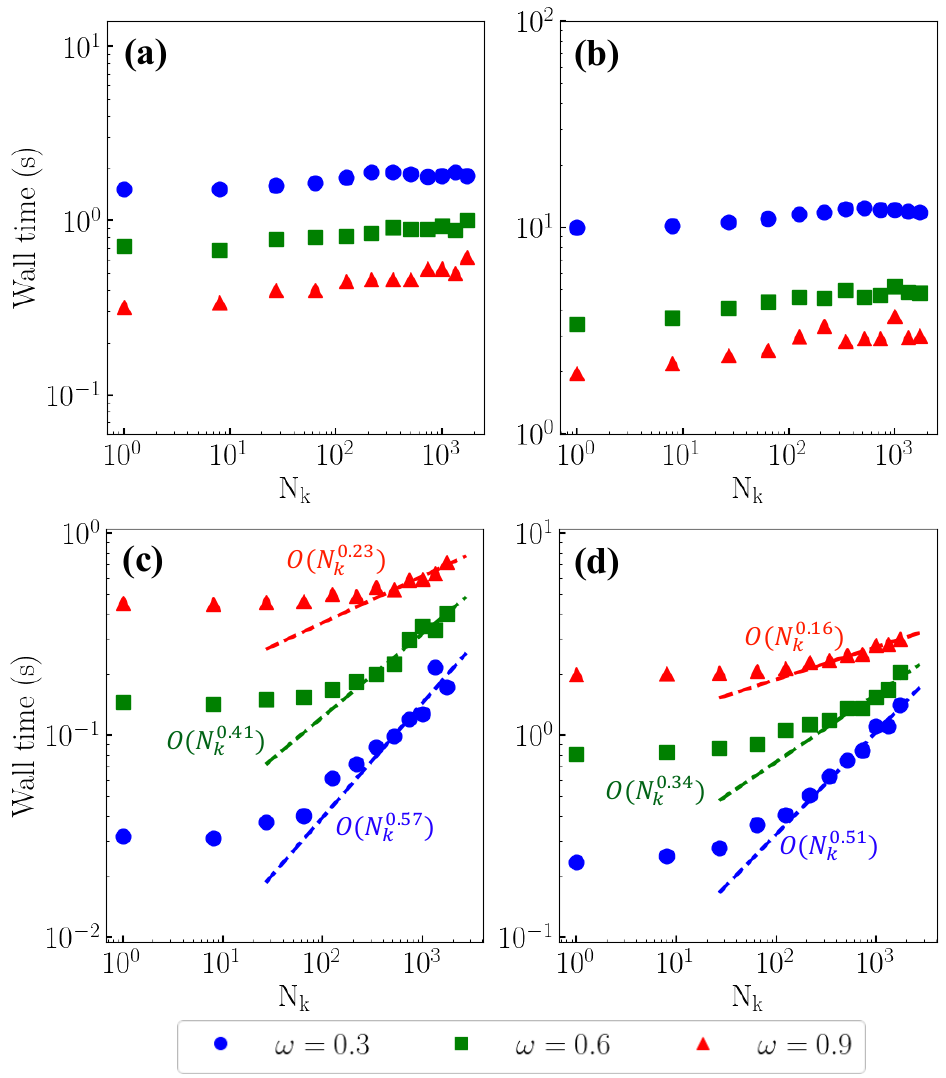}
    \caption{Initialization time at various range-separated parameter values $\omega$ for Si crystal with \textbf{(a)} cc-pVDZ and \textbf{(b)} cc-pVQZ basis set. 
    The initialization time consists of calculating the three-center integral and the two-center integral.
    Time per iteration for various range-separated parameter values is also included for the Si crystal with \textbf{(c)} cc-pVDZ and \textbf{(d)} cc-pVQZ basis sets.
    The number of sampled $\mathbf k$-mesh is from $1\times1\times1$ to $12\times12\times12$.}
    \label{fig:si crystal wall time for dz and qz}
\end{figure}

\begin{figure}[H]
    \centering
    \includegraphics[width=0.6\linewidth]{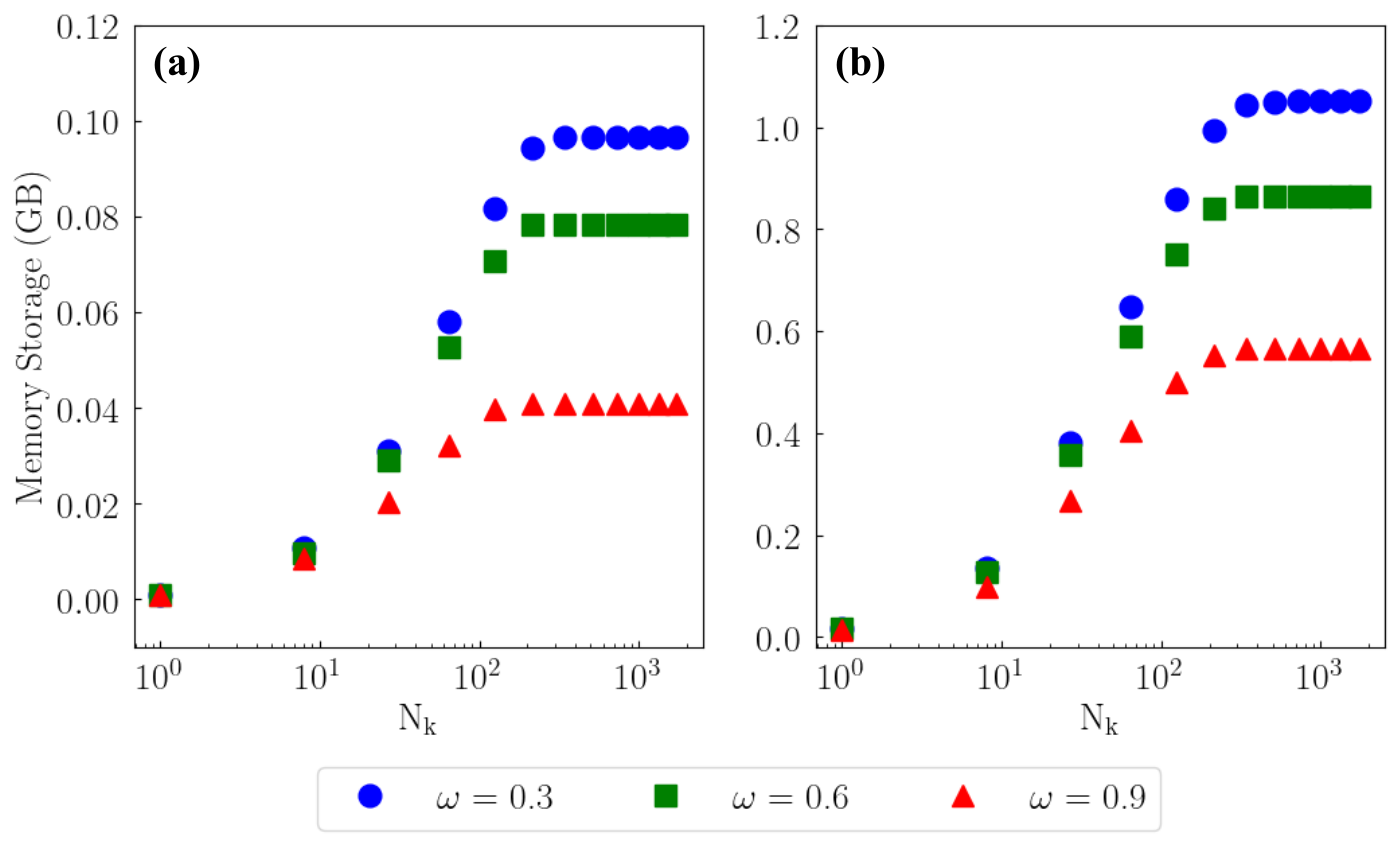}
    \caption{Memory requirement for storing two-electron three-center short-range integrals of Si crystal with \textbf{(a)} cc-pVDZ and \textbf{(b)} cc-pVQZ basis set. The number of sampled $\mathbf k$-mesh is from $1\times1\times1$ to $12\times12\times12$.}
    \label{fig:si dz and qz storage}
\end{figure}

We present the wall time and storage requirements as a function of $N_k$ for the Si crystal with cc-pVDZ and cc-pVQZ basis sets in \cref{fig:si crystal wall time for dz and qz} and \cref{fig:si dz and qz storage}. 
These results capture the same qualitative features as the Si crystal calculation using the cc-pVTZ basis set in the main text.

\newpage
\section{Acceleration of Coulomb matrix construction} \label{sec:J build acceleration for grid based}

In this section, we present an efficient algorithm to construct the Coulomb matrix given the planewave-fitted potential $V(\mathbf{G})$. Relevant coefficients to transform the Coulomb matrix in \cref{alg: J matrix transformation} can be obtained by reversing the sequence of transformations for the density matrix in the main text.

\begin{algorithm} [H]
\caption{Evaluate $J_{\mu\nu}^{\mathbf{0}\mathbf{n}}$ given $V(\mathbf{G})$} \label{alg: non-ortho J eval}
\begin{algorithmic}
    \FOR{primitive shell pair in $| \mu \mathbf{0} \nu \mathbf{n})$}
        \STATE Evaluate $I_{l_{G_i}}(\mathbf{G}_i)$, $I_{l_{G_j}}(\mathbf{G}_j)$, $I_{l_{G_k}}(\mathbf{G}_k )$ recursively.
        \STATE Evaluate $E(\mathbf{G}_i, \mathbf{G}_j)$, $E(\mathbf{G}_j, \mathbf{G}_k)$, $E(\mathbf{G}_k, \mathbf{G}_i)$ recursively.
        \STATE Initialize $J_{l_{G_i}, l_{G_j}, l_{G_k}}$.
        \FOR{$G_k$ in $\{\mathbf{G}_k\}$}
            \STATE Initialize $J_{l_{G_i}, l_{G_j}}(G_k)$.
            \FOR{$G_j$ in $\{\mathbf{G}_j\}$}
                \STATE Initialize $J_{l_{G_i}}(G_j, G_k)$.
                \FOR{$G_i$ in $\{\mathbf{G}_i\}$}
                    \STATE $I_{l_{G_i}}(G_i, G_j, G_k) \leftarrow I_{l_{G_i}}(G_i) \cdot E(G_i, G_j) \cdot E(G_k, G_i)$
                    \STATE $J_{l_{G_i}}(G_j, G_k) \mathrel{+} = V(G_i,G_j,G_k) \cdot I_{l_{G_i}}(l_{G_i}, G_i, G_j, G_k)$  
                \ENDFOR
                \STATE $I_{l_{G_j}}(G_j, G_k) \leftarrow I_{l_{G_j}}(G_j) \cdot E(G_j, G_k)$
                \STATE $J_{l_{G_i}, l_{G_j}}(G_k) \mathrel{+} = I_{l_{G_j}}(G_j, G_k) \cdot J_{l_{G_i}}(G_j, G_k)$
            \ENDFOR
            \STATE $J_{l_{G_i}, l_{G_j}, l_{G_k}} \mathrel{+} = J_{l_{G_i}, l_{G_j}}(G_k) \cdot I_{l_{G_k}}(G_k)$
        \ENDFOR
        
        \STATE Use~\cref{alg: J matrix transformation} to transform $J_{l_{G_i}, l_{G_j}, l_{G_k}}$ to $J_{\mu\nu}^{\mathbf{0}\mathbf{n}}$.
    \ENDFOR
\end{algorithmic}

\end{algorithm}

\begin{algorithm} [H]

\caption{Transformation to BvK Coulomb matrix $J_{\mu\nu}^{\mathbf{0}\mathbf{n}}$} \label{alg: J matrix transformation}
\begin{algorithmic}
    \FOR{primitive shell pair in $|\mu \mathbf{0}\nu \mathbf{n})$}
        \STATE Transform $J_{l_{G_i}, l_{G_j}, l_{G_k}}$ to $J_{l_{G_x}, l_{G_y}, l_{G_z}}$.
        \STATE Transform $J_{l_{G_x}, l_{G_y}, l_{G_z}}$ to $J_{l_{x_p}, l_{y_p}, l_{z_p}}$.
        \STATE Transform $J_{l_{x_p}, l_{y_p}, l_{z_p}}$ to $J_{l_{x_a}, l_{y_a}, l_{z_a}, l_{x_b}, l_{y_b}, l_{z_b}}$.
        \STATE Fold $J_{l_{x_a}, l_{y_a}, l_{z_a}, l_{x_b}, l_{y_b}, l_{z_b}}$ into 3D Cartesian GTOs and transform to spherical harmonic Gaussian basis.
    \ENDFOR
\end{algorithmic}

\end{algorithm}

\section{Periodic exchange-correlation calculation} \label{sec:XC details}

The basic ingredients for evaluating all-electron exchange-correlation contribution are an atom-centered grid and a smooth space partition function $p_{A, \mathbf{n}}(\mathbf{r})$. \cite{becke1988multicenter} 
The exchange-correlation energy in periodic calculations is given by
\begin{equation} \label{eq:XC energy}
\begin{aligned}
    E_{\text{XC}} &= \int_{\Omega_0} \mathrm{d} \mathbf{r} \cdot \varepsilon_{\text{XC}}(\mathbf{r}) \\
                  &= \sum_{\mathbf{n}} \sum_{A} \int_{\Omega_0} \mathrm{d} \mathbf{r} \cdot p_{A, \mathbf{n}} (\mathbf{r}) \varepsilon_{\text{XC}}(\mathbf{r}) \\
                  &= \sum_{A} \int \mathrm{d} \mathbf{r} \cdot p_{A, \mathbf{0}} (\mathbf{r}) \varepsilon_{\text{XC}}(\mathbf{r}),
\end{aligned}
\end{equation}
where $\varepsilon_{\text{XC}}(\mathbf{r} + \mathbf{R}_\mathbf{n}) = \varepsilon_{\text{XC}}(\mathbf{r})$ and $p_{A, \mathbf{n}} (\mathbf{r} - \mathbf{R}_{\mathbf{m}}) = p_{A, \mathbf{n} + \mathbf{m}}(\mathbf{r})$.
The normalized partition function is given as 
\begin{equation} \label{eq:normalized partition function}
    p_{A, \mathbf{0}}(\mathbf{r}) = \frac{w_{A,\mathbf{0}} (\mathbf{r})}{ \sum_{B} \sum_{\mathbf{m}} w_{B,\mathbf{m}}(\mathbf{r})},
\end{equation}
where $w_{A,\mathbf{0}}(\mathbf{r})$ is the unnormalized partition function.
Instead of the typical unnormalized partition function that consists of products of smooth switching functions $\prod_{ B\mathbf{m} \neq A\mathbf{0}} s(A\mathbf{0}, B\mathbf{m}, \mathbf{r})$ as in the molecular calculation, 
we use the unnormalized partition function that has been implemented in the Amsterdam Density Functional Program Suite,  \cite{franchini_becke_2013}
\begin{equation} \label{eq:ADFPS unnormalized partition function}
w_{A,\mathbf{n}}(\mathbf{r}) = \eta_A \cdot \frac{e^{-2 |\mathbf{r} - \mathbf{R}_A + \mathbf{R}_{\mathbf{n}}|}}{|\mathbf{r} - \mathbf{R}_A + \mathbf{R}_{\mathbf{n}}|^3},
\end{equation}
where $\mathbf{R}_A$ is the atomic coordinate in the central unitcell, and $\eta_{A}$ is the size adjustment parameter. $\eta_{A}$ is $0.3$ for hydrogen and $1$ for all other elements. 
This unnormalized partition function, $w_{A,\mathbf{n}}(\mathbf{r})$, only depends on the distance between point $\mathbf{r}$ and the nuclei $A$ in cell $\mathbf{n}$.
Therefore, for a given grid point $\mathbf{r}$, we only need to loop through the other centers in the central unit cell and their images once to obtain the Becke weight $p_{A, \mathbf{0}}(\mathbf{r})$.
In our implementation, the number of significant images in \cref{eq:ADFPS unnormalized partition function} is determined by the radius of the most diffuse shell in the basis set.
If $\mathbf{r}$ is too close to the center $B \mathbf{m}$, the distance $|\mathbf{r} - \mathbf{R}_B + \mathbf{R}_{\mathbf{m}}|$ is set to $10^{-4}$.

\section{Auxiliary charge density integration} \label{sec:Auxiliary charge integration}

We briefly present equations for evaluating the charge of auxiliary basis function with angular momentum triplet $(l_x, l_y, l_z)$. The integral reads
\begin{equation}
    \int \mathrm{d} \ra \cdot \phi_{p}(l_x, l_y, l_z, \ra) = N_p \int dx \cdot x_p^{l_x} e^{-\alpha x_p^2} \int dy \cdot y_p^{l_y} e^{-\alpha y_p^2} \int dz \cdot z_p^{l_z} e^{-\alpha z_p^2},
\end{equation}
and each component is 
\begin{equation} \label{eq:aux charge integration}
    \int dx \cdot x_p^{l_x} e^{-\alpha x_p^2} = \left( \frac{\pi}{\alpha} \right)^{1/2}  \frac{(l_x - 1)!!}{(2\alpha)^{l_x/2}}.
\end{equation}
The double factorial follows $(l_x - 1)!! = (l_x - 1) \cdot (l_x - 3) \cdots 1$.

\section{Global Gaussian density fitting for short-range Coulomb matrix} \label{sec:J-build SR derivation}

As noted in the main text, we use density fitting to accelerate the short-range component. In this section, we derive relevant equations for short-range integral density fitting.
We fit the diagonal element of the pair density with a Gaussian function as
\begin{equation} \label{eq:density fitting}
    \rho_{\mu\nu}^{\ka\ka}(\ra) = \sum_{P} C_{\mu\nu P}^{\ka} \cdot \phi_{P}^{\tilde{\mathbf{0}}}(\ra),
\end{equation}
and the fitting coefficient can be obtained by solving the linear equation using the short-range metric
\begin{equation} \label{eq:metric equation}
    (\phi_Q^{\tilde{\mathbf{0}}} | \rho_{\mu\nu}^{\ka\ka})_{\omega} = \sum_{P} C_{\mu\nu P}^{\ka} (\phi_Q^{\tilde{\mathbf{0}}} |\phi_P^{\tilde{\mathbf{0}}})_{\omega},
\end{equation}
where we note that the $\mathbf{G} = \mathbf{0}$, which corresponds to the monopole-monopole interaction, has been removed. These integrals are defined as
\begin{align}
    (\phi_Q^{\tilde{\mathbf{0}}} | \rho_{\mu\nu}^{\ka\ka})_{\omega} &= \iint \mathrm{d} \ra_1 \mathrm{d} \ra_2 \phi_Q^{\tilde{\mathbf{0}}}(\ra_1) \frac{\text{erfc}(\omega |\ra_{12}|)}{|\ra_{12}|}  \rho_{\mu\nu}^{\mathbf{k}}(\ra_2) - \frac{\pi}{\Omega \omega^2} Q_Q Q_{\mu\nu}^{\ka} , \\
    (\phi_Q^{\tilde{\mathbf{0}}} |\phi_P^{\tilde{\mathbf{0}}})_{\omega} &= \iint \mathrm{d} \ra_1 \mathrm{d} \ra_2 \phi_Q^{\tilde{\mathbf{0}}}(\ra_1) \frac{\text{erfc}(\omega |\ra_{12}|)}{|\ra_{12}|} \phi_P^{\tilde{\mathbf{0}}}(\ra_2) - \frac{\pi}{\Omega \omega^2} Q_Q Q_P,
\end{align} 
and $\mathbf{0}$, $\tilde{\mathbf{0}}$ refer to the central unit cell and $\Gamma$ point, respectively.
Furthermore, we have the following relation for the charge of the pair density and the auxiliary basis
\begin{equation} \label{eq:charge equation through fitting}
    \sum_{\mathbf{n}} e^{i \ka \R_n}\int \mathrm{d} \ra \rho_{\mu\nu}^{\mathbf{0} \mathbf{n}}(\ra) = \sum_{P} C_{\mu\nu P}^{\ka} \int d \ra \phi_{P}^{\mathbf{0}}(\ra) = \sum_{P} C_{\mu\nu P}^{\ka} \cdot Q_P
\end{equation}
where integration within the central cell is carried out for~\cref{eq:density fitting} and then translated to integration over all space through lattice summation.

We now transform the short-range Coulomb matrix,
\begin{align} \label{eq:Transforming J}
    [J_{\mu\nu}^{\mathbf{k}}]_{SR} &= \sum_{\mathbf{k'}} \sum_{\lambda \sigma} \left[ (\rho_{\mu\nu}^{\mathbf{k} \mathbf{k}} | \rho_{\lambda \sigma}^{\mathbf{k'} \mathbf{k'}})_{\omega} - \frac{\pi}{\Omega \omega^2} Q_{\mu\nu}^{\ka} Q_{\lambda \sigma}^{\ka'} \right] \cdot D_{\lambda \sigma}^{\mathbf{k'}}, \\
    &= \sum_{\ka'} \sum_{PQ} \sum_{\lambda \sigma} C_{\mu\nu P}^{\ka} \left[ \iint \mathrm{d} \ra_1 \mathrm{d} \ra_2 \phi_P^{\tilde{\mathbf{0}}}(\ra_1) \frac{\text{erfc}(\omega |\ra_{12}|)}{|\ra_{12}|} \phi_Q^{\tilde{\mathbf{0}}}(\ra_2) - \frac{\pi}{\Omega \omega^2} Q_P Q_Q \right] C_{\lambda \sigma Q}^{\ka'} \cdot  D_{\lambda \sigma}^{\mathbf{k'}} \\
    &= \sum_{\ka'} \sum_{PQ} \sum_{\lambda \sigma} C_{\mu\nu P}^{\ka} \cdot (\phi_Q^{\tilde{\mathbf{0}}} |\phi_P^{\tilde{\mathbf{0}}})_{\omega} \cdot C_{\lambda \sigma Q}^{\ka'} \cdot  D_{\lambda \sigma}^{\mathbf{k'}} \\
    &= \sum_{\ka'} \sum_{PQ} \sum_{\lambda \sigma} (\rho_{\mu\nu}^{\ka\ka} | \phi_Q^{\tilde{\mathbf{0}}})_{\omega} \cdot (\phi_Q^{\tilde{\mathbf{0}}} |\phi_P^{\tilde{\mathbf{0}}})_{\omega}^{-1} \cdot (\phi_P^{\tilde{\mathbf{0}}} | \rho_{\lambda \sigma}^{\ka'\ka'})_{\omega} \cdot  D_{\lambda \sigma}^{\mathbf{k'}},
\end{align}
where the last equality comes from taking the inverse of ~\cref{eq:metric equation}. The diagonal pair density can then be transformed into the BvK representation to obtain the expression for the Coulomb matrix in the main text.

\end{document}